\documentclass[sigplan,screen]{acmart}

\usepackage{enumitem}
\usepackage{url}
\usepackage{xspace}
\usepackage{algorithm}
\usepackage{algorithmic}
\usepackage{xcolor}
\usepackage{subcaption}
\usepackage{listings}
\usepackage{xcolor}
\usepackage[absolute]{textpos}

\newcommand{\FUNCTION}[2]{%
  \STATE \textbf{function} \textsc{#1}(#2)%
  \begin{ALC@g}%
}
\newcommand{\ENDFUNCTION}{%
  \end{ALC@g}%
  \STATE \textbf{end function}%
}

\newcommand{\STATEX}{\item[]}

\definecolor{mybg}{HTML}{E0ECF4}

\NewDocumentEnvironment{mybox}{O{0.48\textwidth} +b}
{%
  \par
  \setlength{\fboxsep}{6pt}%
  \setlength{\fboxrule}{0.4pt}%
  \noindent
  \fcolorbox{black}{mybg}{%
    \begin{minipage}{\dimexpr#1-2\fboxsep-2\fboxrule\relax}
      #2
    \end{minipage}%
  }%
  \par
}
{}

\newcommand{\THISWORK}{{\fontfamily{lmss}\selectfont
KAIROS}\xspace}

\author{Yichao Yuan}
\affiliation{%
  \institution{University of Illinois Urbana-Champaign}
  \city{Urbana}
  \state{IL}
  \country{USA}
}
\email{yichaoy2@illinois.edu}

\author{Mosharaf Chowdhury}
\affiliation{%
  \institution{University of Michigan}
  \city{Ann Arbor}
  \state{MI}
  \country{USA}
}
\email{mosharaf@umich.edu}

\author{Nishil Talati}
\affiliation{%
  \institution{University of Illinois Urbana-Champaign}
  \city{Urbana}
  \state{IL}
  \country{USA}
}
\email{nishil@illinois.edu}

\begin{document}
\begin{textblock}{14}(5,0.5)
{\normalsize \normalfont \textit{To appear in the Proceedings of ASPLOS 2027.} }
\end{textblock}

\title{\THISWORK: Stateful, Context-Aware Power-Efficient Agentic Inference Serving}

\begin{abstract}
Power has become a central bottleneck for AI inference.
This problem is becoming more urgent as agentic AI emerges as a major workload class, yet prior power-management techniques focus almost entirely on single-turn LLM serving.
Our analysis shows that agentic serving behaves fundamentally differently: each request carries long-lived context that evolves across tool-interleaved turns, and lowering GPU frequency can push the system into a thrashing regime where memory pressure sharply worsens both performance and power efficiency.
These observations show that power optimization for agentic serving requires rethinking.

We present \THISWORK, a context-aware power optimization system for agentic AI serving.
\THISWORK\ uses \textit{agent context as a first-class control signal} to jointly manage GPU frequency, per-instance concurrency, and multi-instance request placement.
This enables \THISWORK\ to save power when memory headroom exists while avoiding thrashing and preserving performance targets.
At a high level, \THISWORK\ tracks requests at agent granularity, adapts local control to context growth and agent progress, and routes agents across instances to jointly improve power efficiency and memory stability.
Evaluated across diverse software and data engineering agentic tasks, \THISWORK\ achieves an average of 27\% (up to 39.8\%) power reduction while meeting the performance targets.
\end{abstract}



\settopmatter{printacmref=false}

\setcopyright{none}
\renewcommand{\footnotetextcopyrightpermission}[1]{}

\maketitle 

\section{Introduction} \label{section:introduction}

Power is a \textit{first-order constraint} for AI systems because AI data center demand has far outpaced the growth of available energy and grid capacity~\cite{iea2025energyai,iea2026electricitygrids,mural2026aidatacentersgrid}.
Reducing power is therefore critical not only for sustainability, but also for improving throughput within a fixed facility power envelope by enabling more usable compute capacity~\cite{stojkovic2025dynamollm,qiu2024muserve,stojkovic2025tapas}.
This urgency is likely to grow because recent trends~\cite{asgar2025agentic,wadlom2026helium} show \textit{agentic AI} as a major emerging inference workload, and serving agentic AI requires orders of magnitude higher power than traditional LLM serving~\cite{kim2026cost}.
This is because agentic AI replaces a single inference with many tool-interleaved, stateful LLM invocations, often requiring dozens of model calls per request and thereby driving much higher computational cost.
Therefore, \textit{improving the power consumption of agentic AI serving is both urgent and critical.}

A large body of prior work has explored reducing power and energy cost of \textit{single-turn LLM serving}.
One major class of techniques uses dynamic frequency and power control.
For example, DVFS-based runtimes that tune GPU frequency online using signals such as workload slack, phase behavior, or workload state~\cite{stojkovic2025dynamollm,kakolyris2024throttllem,greenllm_dac,basit2026biscale,yu2025voltanallmfeedbackdrivenfrequencycontrol,wang2025usinganalytical,qiu2024muserve,spaan2026kerneldvfs}. 
A second class improves efficiency through system-level scheduling and resource management, such as hardware-aware placement, heterogeneous scheduling, and power-aware cluster orchestration~\cite{wilkins2024hybrid,li2025ecoserve,stojkovic2025tapas}.
However, these techniques are designed for \textit{stateless} LLM serving, where requests are short-lived and largely independent.
This raises a fundamental question: \textit{can power saving techniques designed for single-turn LLM serving be applied directly to agentic AI serving?}

To answer this question, we conduct a detailed characterization of agentic AI workloads from the perspective of an LLM serving system.
Specifically, we focus on ReAct-style agents~\cite{yao2023react} because prior work has shown that they lie near/on the intelligence--cost Pareto frontier~\cite{kim2026cost}.
We find that, unlike single-turn serving, agentic workloads maintain \textit{long-lived state} for each request through a \textit{context cache}~\cite{kwon2023pagedattention} that grows across tool-interleaved turns and varies widely across agents.
Our study further shows that this context growth is \textit{highly dynamic}: it varies substantially over time within and across agents, and the distributions of agent lifetimes, turn counts, and growth rates all exhibit long tails.
This makes memory pressure difficult to predict and control, and creates a fundamental tension for power management: lowering GPU frequency saves power, but it also slows execution, and in agentic workloads that slowdown accumulates across turns and can directly hurt per-agent progress.

Our analysis also uncovers a new power-management phenomenon: lowering GPU frequency beyond a certain value can drive the system into a \textit{thrashing regime}, where accumulated context exceeds GPU memory capacity.
As a result, the context needs to be recomputed~\cite{kang2026thunderagent,li2025continuum} or loaded from slower memory tiers~\cite{cheng2025lmcache} that degrade throughput, latency, and power efficiency.
This strong coupling between power and memory stability shows that agentic serving is not merely a scaled-up version of single-turn LLM serving, but a fundamentally different workload whose stateful and dynamic behavior reshapes the power-performance trade-off.
As a result, \textit{we must rethink the problem of power optimization for agentic AI} around context growth, memory stability, and thrashing-aware control.

In this paper, we present \THISWORK, a context-aware power management system for agentic AI serving that reduces power consumption while maintaining a target performance SLO.
Our goal is to minimize system power without significantly sacrificing end-to-end agent progress, despite the highly dynamic and stateful nature of agentic workloads.
To achieve this, \THISWORK\ is designed around three key principles: (1) dynamically adjusting GPU frequency based on context pressure to safely trade performance for power savings, (2) regulating concurrency to prevent excessive context growth that can destabilize the system, and (3) scaling and routing requests across multiple serving instances in a power-aware manner to consolidate load at low demand and spread load under high context pressure.

To realize these goals, \THISWORK\ introduces a unified runtime that combines lightweight request tracking, per-instance control, and global routing. 
First, it \textit{tracks requests at the agent granularity}, enabling the system to reason about long-lived, multi-turn workflows rather than isolated LLM calls.
Each serving instance is then equipped with a \textit{context-aware controller} that monitors context growth and system load to adapt GPU frequency and admission decisions.
This allows \THISWORK\ to save power when capacity exists while avoid thrashing and preserving performance.
Finally, a \textit{global router} places agents across serving instances based on context and load, consolidating under light demand and spreading under high context pressure.
Together, these components use \textit{agent context as a first-class signal} for power-efficient and stable agentic AI serving.

We evaluate \THISWORK\ on  NVIDIA H100 GPUs that run vLLM~\cite{kwon2023pagedattention} across three datasets, including SWE-bench Verified~\cite{chowdhury2024swebenchverified}, DABStep~\cite{egg2025dabstep}, and Terminal-Bench 2.0~\cite{merrill2026terminalbench}, and two agent types: mini-swe-agent~\cite{yang2024sweagent} and Terminus-2~\cite{merrill2026terminalbench}.
Unlike single-turn LLM serving that relies on TTFT/TBT, we adopt a throughput-based SLO since agentic execution is best measured by the rate of task progress (\S\ref{section:problem_formulation}).
\THISWORK\ reduces average GPU power by \textbf{27\%} (up to \textbf{39.8\%}) for a single serving instance and \textbf{46.3\%} for multi-instance serving while maintaining per-agent SLO of P5 throughput 20 tokens/s.
\THISWORK\ show effectiveness in avoiding thrashing, demonstrating stable operation under dynamic context growth.

To the best of our knowledge, \textbf{\THISWORK\ is the first work to optimize power for agentic inference serving}.
Our key contributions are:
\begin{itemize}[leftmargin=*]
    \item A detailed characterization of agentic inference serving that reveals why power optimization is fundamentally different and harder compared to single-turn LLM serving.
    \item A context-aware control mechanism that jointly manages GPU frequency and concurrency using context growth, memory pressure, and agent progress to reduce power while avoiding thrashing and preserving SLO attainment.
    \item \THISWORK: an end-to-end system for agentic AI serving that reduces power by 27\%, maintaining performance targets.
\end{itemize}
\section{Background: Agentic Inference Serving and Power/Energy Efficiency} \label{section:background}

This section provides brief background on the agentic inference serving workload, how is it different than single-turn LLM inference, and the need for power efficiency.

\subsection{Agentic AI Inference Serving}
Agentic inference differs from single-turn LLM inference in that the model is invoked \textit{repeatedly in a closed-loop process} that alternates between reasoning, acting, and observing feedback, rather than a single prompt-response exchange.
A few examples of existing agent designs include ReAct~\cite{yao2023react}, Reflexion~\cite{shinn2023reflexion}, and LATS~\cite{zhou2024lats}, which differ mainly in how they organize control flow around the model. 
A recent characterization study~\cite{kim2026cost} found that ReAct often lies on or near the accuracy--cost Pareto frontier. 
Therefore, we focus our discussion on \textbf{agents following ReAct paradigm} because they are widely adopted and effective across diverse tasks including multi-hop question answering~\cite{shridhar2021alfworld}, web shopping~\cite{yao2022webshop}, data engineering~\cite{egg2025dabstep}, and software engineering~\cite{jimenez2024swebench}.
As we will demonstrate in evaluation (\S~8.2), our techniques are generally applicable to advanced production-grade agents like Claude Code~\cite{anthropic2025claudecode}.

\begin{figure}[t]
    \centering
    \includegraphics[width=\linewidth]{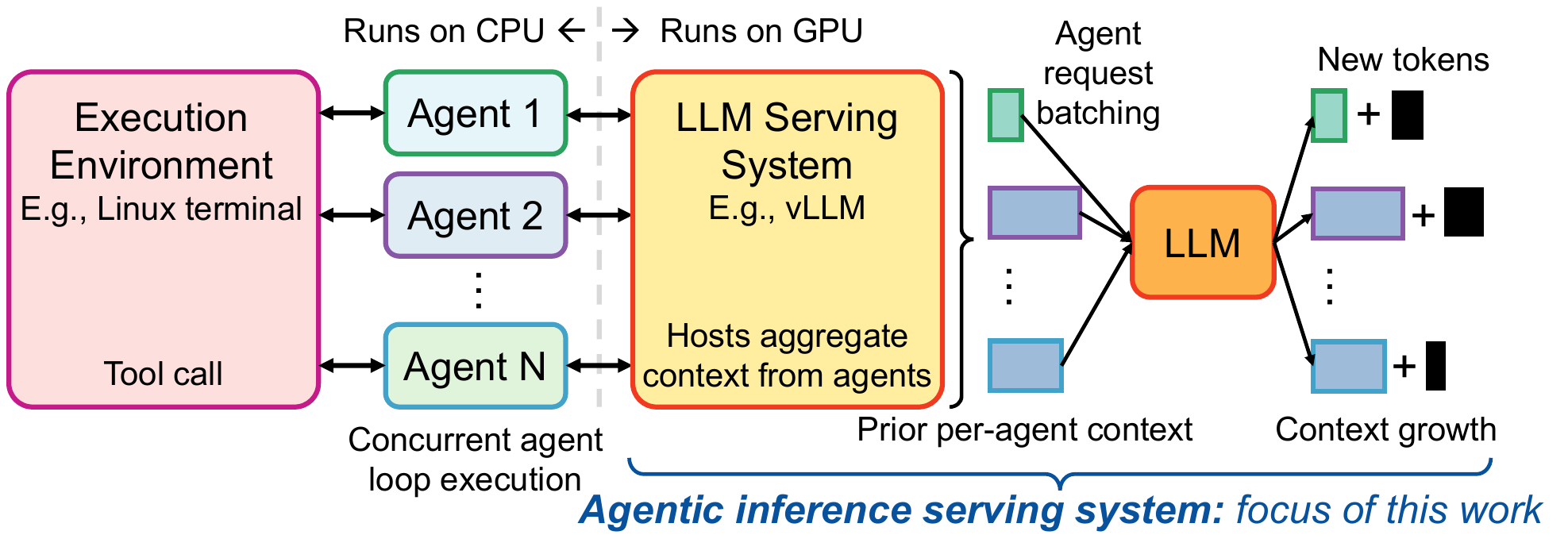}
    \caption{Example of ReAct agent workflow: multiple concurrent agents perform multi-turn conversations between execution environment and the LLM serving system.
    }
\label{fig:agent_background}
\end{figure}

As shown in Figure~\ref{fig:agent_background}, a ReAct agent repeatedly sends prompts and tool outputs to an LLM server, such as vLLM~\cite{kwon2023pagedattention}, executes the returned action in an execution environment, and feeds the resulting observation into the next turn.
When many agents run in parallel, the serving system must handle many interleaved prefill and decode phases whose \textit{per-agent contexts persist and grows} across turns rather than disappearing after one response.
In systems such as vLLM, previously processed tokens are retained in the \textit{prefix/context cache}~\cite{kwon2023pagedattention} so later turns can resume from accumulated history instead of recomputing the entire conversation from scratch.
Once the agent completes a task, the entire context becomes useless, freeing up a large GPU memory capacity.
This makes agentic inference serving fundamentally \textbf{\textit{stateful}}, because each new turn depends on prior context.

A concrete example of a ReAct agent is SWE-Bench~\cite{jimenez2024swebench} that resolves GitHub issues, where an agent iteratively inspects code, runs commands and tests, revises its hypothesis, edits files, and re-queries the model until it produces a valid patch.
Tokens from each turn are retained in a context cache, making serving inherently \textit{stateful}, while actions in the environment are guided by the language model.

\subsection{An Urgent Need for Power/Energy Efficiency}
AI data centers face severe energy and power pressure, with inference as a major contributor and modern deployments operating at hundreds of megawatts to gigawatts~\cite{luccioni2024powerhungry,kim2026cost,stojkovic2025dynamollm}.
This paper focuses on \textit{power} because data centers are increasingly constrained by available watts; reducing per-server power can increase deployable compute capacity within a fixed facility power envelope~\cite{belfer2025aidatacentersgrid}. 
This issue is becoming more urgent as recent work~\cite{asgar2025agentic,wadlom2026helium} points to agentic AI as an emerging workload class, and prior characterization~\cite{kim2026cost} shows that agentic inference consumes two to three orders of magnitude higher energy than single-turn LLM.
This gap arises because agentic serving repeatedly alternates prefill, decode, and tool phases over many turns while the serving system retain prior tokens in the context cache, causing GPU memory usage to grow unevenly and intensify both capacity and bandwidth pressure. 
Although agentic execution uses CPUs for orchestration and tool interaction, the GPU-side LLM serving path remains the dominant power-hungry component, which is the focus of our work (Figure~\ref{fig:agent_background}).



GPU power can be reduced through static power caps, dynamic frequency scaling, or broader thermal/power-aware scheduling \cite{qiu2024muserve,kakolyris2024throttllem,spaan2026kerneldvfs,stojkovic2025tapas}.
In our workload setting, agent context evolves dynamically across turns, making static policies suboptimal fit, motivating DVFS as an online control mechanism.
Because lowering frequency also slows inference down, any such approach must balance power savings against a performance target SLOs.
SLOs in LLM serving are defined using latency metrics:  TTFT and TBT, which are designed to capture user-perceived responsiveness in interactive applications~\cite{agrawal2024sarathiserve}.
In contrast, we argue for a throughput-based SLO for agentic serving with more details in \S\ref{section:problem_formulation}. 


\section{Characterization of Agentic Inference Serving} \label{section:characterization}


\subsection{Experimental setup} \label{subsection:characterization:setup}
While \S\ref{section:methodology} presents a detailed methodology used throughout the paper, this subsection briefly discusses the experimental setup used for our characterization study.
We run agentic AI inference using the Harbor~\cite{Harbor_Framework} framework on a single NVIDIA H100 GPU with context caching enabled~\cite{kwon2023pagedattention}.
Unless otherwise specified, the GPU runs a vLLM server that with a Qwen3-Coder-30B model with SWE-bench Verified~\cite{chowdhury2024swebenchverified} dataset and mini-swe-agent~\cite{yang2024sweagent}.
We use a fixed input request rate of 0.08 agent jobs per second using a Poisson distribution to represent real-world traffic patterns~\cite{mlperf_inference_datacenter}.
Therefore, in the steady state, vLLM serves requests from multiple agents concurrently.
Each agent job will perform multiple conversation turns, resulting in a large batch size.

\subsection{Dynamic Context Growth} \label{subsection:characterization:dynamic_growth}
\begin{figure}[t]
    \centering
    \includegraphics[width=\linewidth]{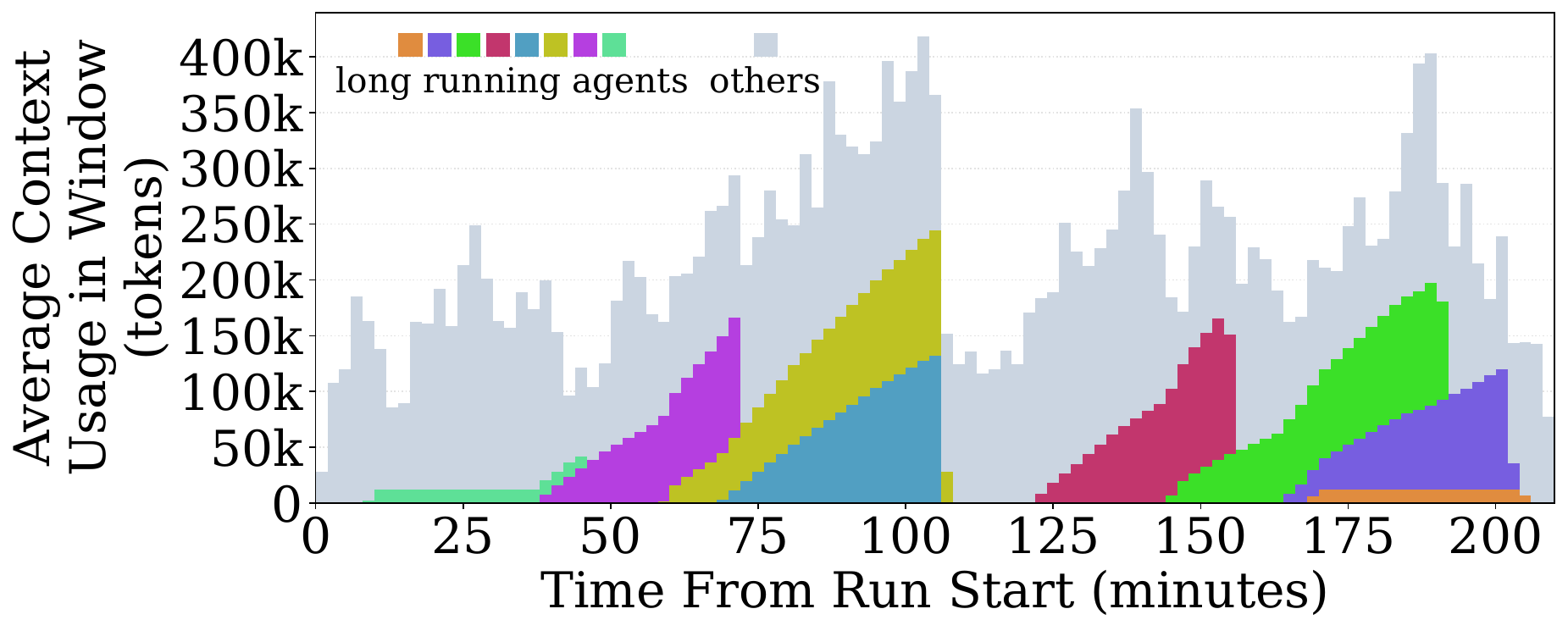}
    \caption{Agent context growth over time for concurrent workloads served by a single vLLM instance; the eight longest contexts are highlighted in distinct colors, with others in gray.}
\label{fig:context_dynamism}
\end{figure}

\begin{figure}[t]
        \centering
        \includegraphics[width=\linewidth]{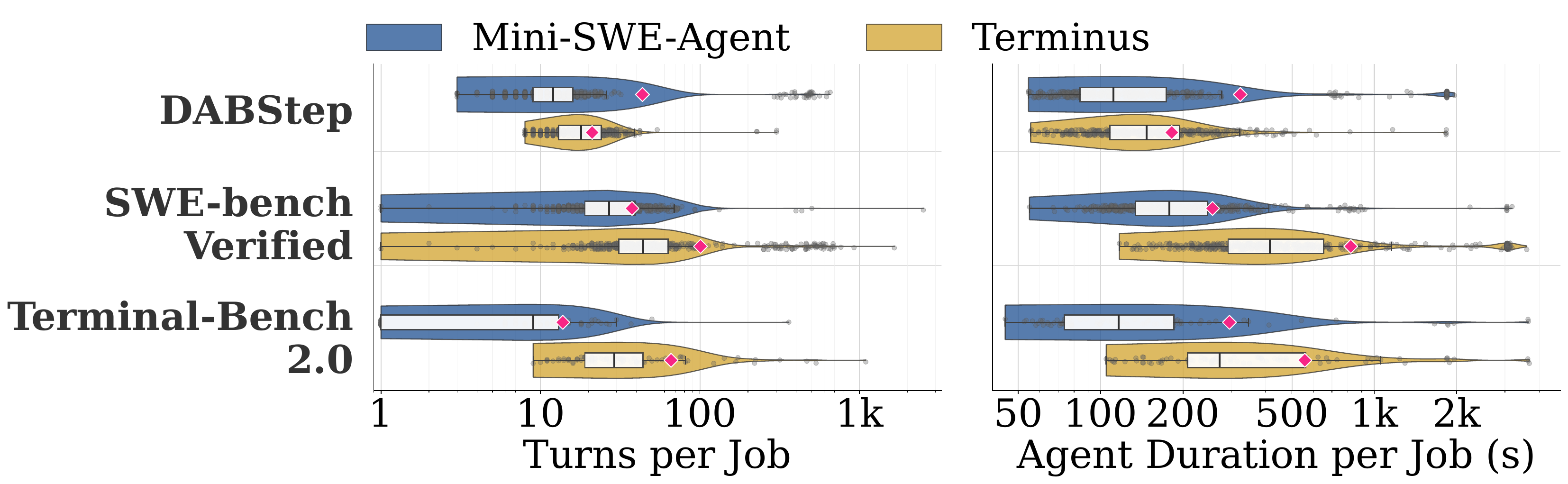}
        \label{fig:turn_count_distribution}
    \caption{Conversation turn count (left) and agent duration (right) log distribution across two agent types and three datasets, showing a high degree of variation.}
    \label{fig:turn_counts}
\end{figure}




Figure~\ref{fig:context_dynamism} illustrates how the context cache capacity of serving concurrent agents evolves over time.
The eight longest contexts are distinguished using different colors, while all remaining contexts are shown in gray to improve the readability of the figure.
Unlike single-turn LLM serving, where workload variability arises mainly from input/output sequence lengths~\cite{yu2022orca,kwon2023pagedattention,agrawal2024sarathiserve}, agentic serving introduces an additional source of variation: long-lived context capacity evolves \textbf{\textit{dynamically}} in three unique ways.

\textbf{First}, the capacity and longevity of each context depends on the \textit{number of conversation turns}.
For example, our setup serves an average of 17 agents in parallel.
The number of conversation turns vary from 1 to 2518, with an average of 37 turns, that depends on the complexity of agent task.
This makes the length of each agent context and its longevity in the serving system a function of the input task, which cannot be determined statically.

To understand this behavior further, Figure~\ref{fig:turn_counts} shows the distribution of number of conversation turns and total time spent by each agent in the LLM serving system.
Due to space limitation, the distribution of maximum context length per agent is shown in Figure~\ref{fig:characterization:context_distribution} (\S\ref{appendix:context_variation}).
Results reveal significant variability in turn counts and agent duration.
Specifically, the heavy-tailed distribution highlights that while many jobs conclude quickly, a non-trivial portion of \textit{long-running} tasks requires thousands of turns.

Because agents have different lifetimes, \textbf{second}, context accumulation across all agents running concurrently is highly uneven over time. 
When an agent completes, its entire context cache is released at once, causing an abrupt drop in the total accumulated context footprint.
\textbf{Third}, the rate of context growth varies largely across multiple agents.
This rate is a function of runtime properties including input and output sequence lengths for each conversation, and the amounts of time spent on LLM and tool calls.
This analysis clearly shows that agentic serving creating highly dynamic and unpredictable memory demands for the serving system.


\subsection{Effect of GPU Frequency Scaling} \label{subsection:characterization:frequency_scaling}
\begin{figure}
    \centering
    \includegraphics[width=\linewidth]{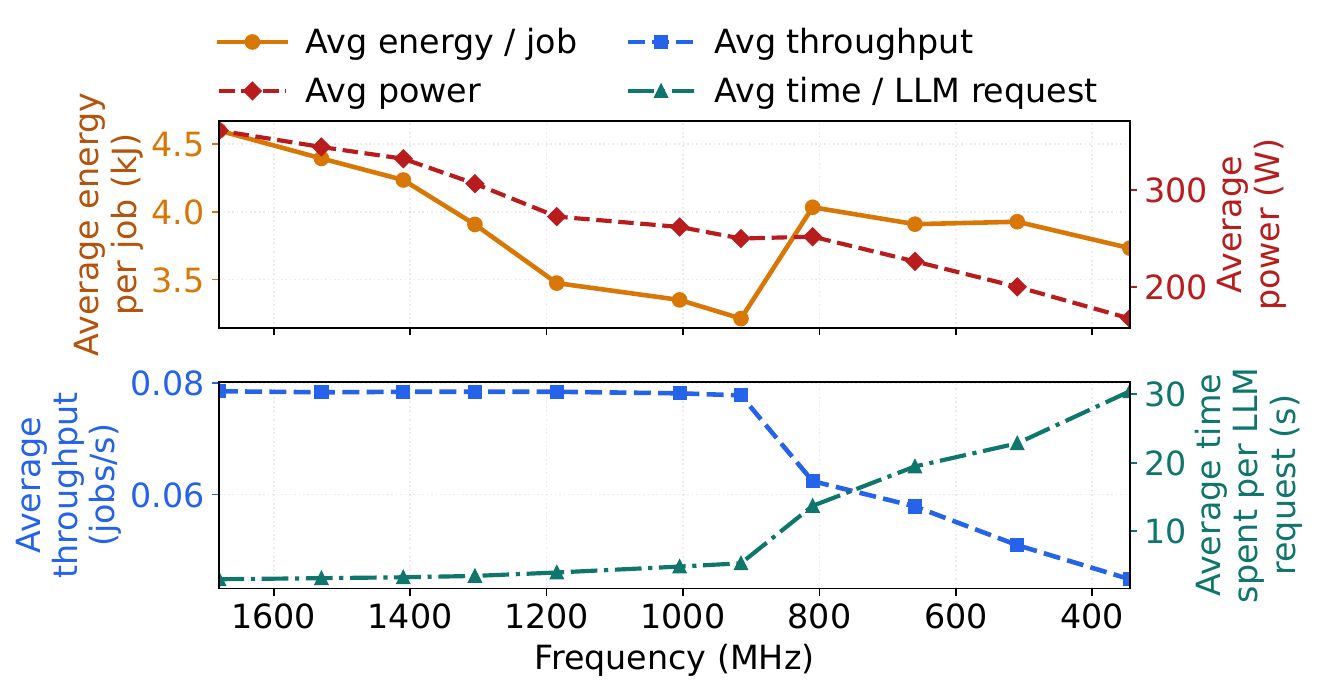}
    \caption{Effect of reducing GPU frequency on energy and power consumption (top), throughput, and  average time per LLM request (bottom) for agentic serving system.}
\label{fig:frequency_scaling_effect}
\end{figure}

Reducing GPU core frequency is a common way to lower power, but it also creates a trade-off with inference performance. 
Prior work has only studied this trade-off in single-turn LLM serving~\cite{chung2026joulesgodiagnosinginference,kakolyris2024throttllem,greenllm_dac,stojkovic2025tapas,stojkovic2025dynamollm,maliakel2025characterizing,yu2025voltanallmfeedbackdrivenfrequencycontrol}.
Figure~\ref{fig:frequency_scaling_effect} shows how GPU frequency affects agentic serving efficiency.
Two observations stand out: (1) reducing frequency from 1680 MHz to 900 MHz lowers energy and power by about 30\% with little impact on throughput or latency, but (2) below 900 MHz this trend reverses, with sharply higher energy, lower throughput, and longer time for each LLM request.

This reversal occurs because lowering frequency slows execution while the request arrival rate remains fixed at 0.08 agents/s (\S\ref{subsection:characterization:setup}).
As service slows, pending work accumulates and the total context footprint of concurrent agents grows until it exceeds GPU memory capacity, pushing the system into a \textbf{\textit{thrashing regime}} where cached context is evicted and previously processed tokens must be \textit{recomputed}.
This extra work raises energy consumption, reduces throughput, and increases latency.
Although memory pressure can also appear in single-shot LLM serving, agentic workloads amplify this effect over multi-turn serving.


\subsection{Deep Dive Into Context Thrashing} \label{subsection:characterization:thrashing_deep_dive}
\begin{figure}[t]
    \centering
    \includegraphics[width=\linewidth]{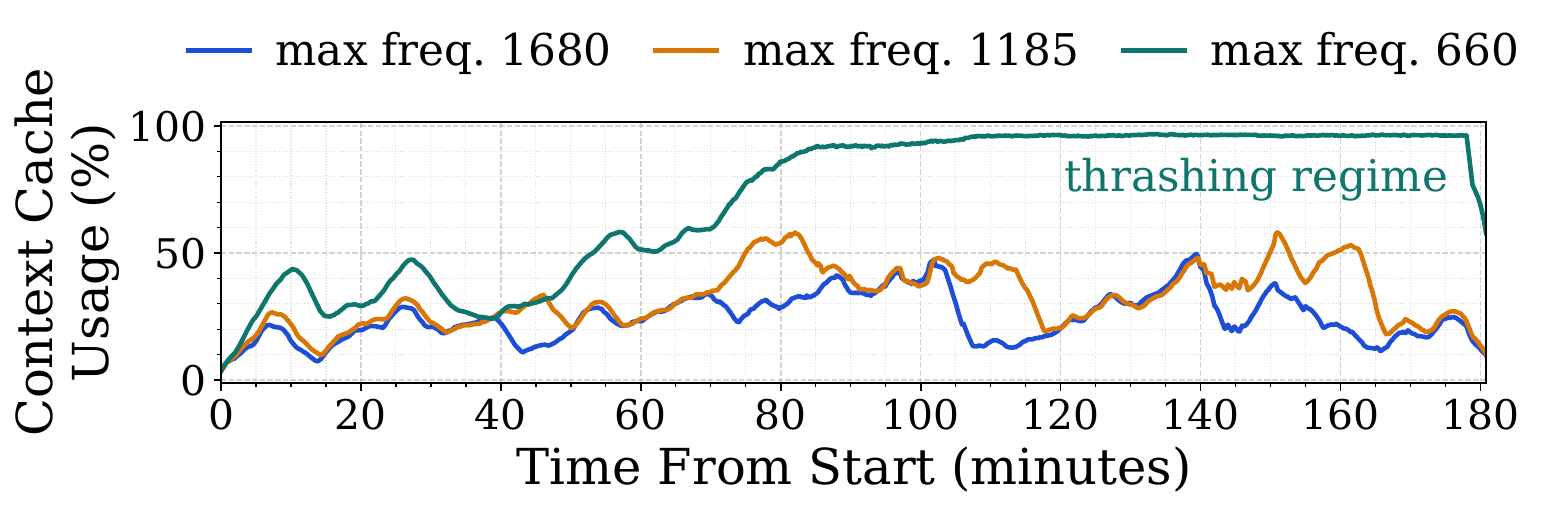}
    \caption{Aggregate context cache growth over time for concurrent agents running at three maximum GPU core frequencies: 1680 MHz, 1185 MHz, and 660 MHz.}
    \label{fig:context_cache_growth_frequency}
\end{figure}
Based on these observations, we classify the serving system into two regimes: \textbf{non-thrashing}, where the aggregate context of concurrent agents fits within available GPU memory, and \textbf{thrashing}, where it exceeds that capacity.
Figure~\ref{fig:context_cache_growth_frequency} shows the transition by plotting aggregate context growth at 1680, 1185, and 660 MHz under a fixed request rate of 0.08 agents/s.
At 1185 MHz, execution slows only slightly relative to 1680 MHz, so tasks still complete fast enough to keep memory pressure stable.
At 660 MHz, however, compute throughput drops enough that accumulated context grows faster than it can be drained, exceeds available memory, pushing the system into a thrashing regime.
This result shows that \textit{frequency-induced compute bottlenecks are a root cause of both energy and performance instability in agentic serving}. 
Appendix~\S\ref{appendix:threashing_performance} further examines thrashing performance under two common fallbacks: recomputation~\cite{kang2026thunderagent,li2025continuum} and LMCache-based offloading~\cite{cheng2025lmcache}.

\subsection{Summary of Unique Challenges} \label{subsection:challenges}
Based on our characterization, we identify three unique \textbf{\underline{C}}hallenges for power saving in agentic inference serving.

\begin{itemize}[leftmargin=*]
    \item \textbf{\underline{C1:} Tension between power saving and performance.} Lowering GPU frequency reduces power, but also slows agent execution, creating a fundamental trade-off. The unique challenge is that the slowdown is \textit{context-dependent and  accumulates across turns}, making it difficult to reduce power without violating a performance target.

    \item \textbf{\underline{C2:} Highly dynamic context demand.} In agentic serving, both the size and the lifetime of each agent's context evolve online as a function of task complexity and tool interaction, making memory demand challenging to predict or provision statically.
    
    \item \textbf{\underline{C3:} Strong coupling between power and memory stability.} Reducing GPU frequency can save power, but slower execution prolongs agent lifetime and increases resident context, which can push the system into a \textit{thrashing regime} that sharply degrades throughput, latency, and power efficiency.
\end{itemize}
These challenges differ fundamentally from single-turn LLM serving, motivating a rethinking of power optimization.
\section{Problem Formulation} \label{section:problem_formulation}

We model an agentic serving system $\mathcal{S}$ in Figure~\ref{fig:agent_background} as a collection of underlying serving instances $\mathcal{M}$.
A serving instance $m \in \mathcal{M}$ is defined as a single unit of an LLM server that either a single or multiple GPUs, depending on the size of the model.
The system processes an incoming stream of agents using a collection of instances.
Let $\mathcal{A}$ denote the set of agents, where each agent $a \in \mathcal{A}$ generates a sequence of requests:
\begin{equation}
a = (r_1, r_2, \dots), \quad r_i = (\Delta p_i, \Delta d_i), \; r_i \in \mathcal{R},
\end{equation}
arriving at a target rate (e.g., agent requests per second),
where $\Delta p_i$ is the number of newly appended prefill tokens (e.g., initial prompt or environment observation) and $\Delta d_i$ is the number of generated decode tokens.

The context (history) after step $i$ is the concatenation of all prior tokens:
\begin{equation}
H_i = (\Delta p_1, \Delta d_1,\Delta  p_2,\Delta  d_2, \dots, \Delta p_i), \quad C_i = |H_i|.
\end{equation}

Each request $r_i$ is processed with end-to-end latency $t_i$, which depends on system conditions and the current context size $C_i$. 
We define the per-agent throughput as:
\begin{equation}
\text{Throughput}(a) = \frac{\sum_i \Delta d_i}{\sum_i t_i},
\end{equation}
and say that an agent $a$ satisfies a throughput-based SLO with target $\tau$ (tokens/s) if $\text{Throughput}(a) \geq \tau$.

While prior work uses latency-based SLOs such as TTFT, TBT, or TTLT~\cite{zhang2026jitserve} (time to last token), we argue that \textit{per-agent throughput} is a more \textit{user-centric} metric that better captures end-to-end agent progress over its full lifecycle.
TTFT and TBT are designed for fine-grained, human-facing generation, whereas agentic serving operates at the task level with multi-turn, autonomous execution.
TTLT is less actionable for online control because it depends on an isolated request-specific latency that is not available in online serving at runtime.
In contrast, throughput directly reflects runtime agent progress under shared execution, making it a more \textit{actionable} SLO for agentic serving.

Let $\mathcal{A}(T)$ denote the set of agents that complete within a time interval $[0, T]$, and let $|\mathcal{A}(T)|$ be its cardinality. 
We define the SLO attainment rate over $[0, T]$ as:
\begin{equation}
\text{SLO-Attainment}(T) = \frac{1}{|\mathcal{A}(T)|} \sum_{a \in \mathcal{A}(T)} \mathbf{1}\big[\text{Throughput}(a) \geq \tau \big].
\end{equation}

We define the total instantaneous system power and the average system power over $[0, T]$ as:
\begin{equation}
P(t) = \sum_{m \in \mathcal{M}} P_m(t), \quad
\bar{P}(T) = \frac{1}{T}\int_0^T P(t)\,dt,
\end{equation}
where $P_m(t)$ denotes the instantaneous power of instance $m$ at time $t$.
Given these definitions, we define an optimization objective as follows.

\begin{mybox}
\textbf{Optimization objective:} Minimize the power consumption $\bar{P}(T)$ while maintaining $\text{SLO-Attainment}(T)$ comparable to a non power-optimized baseline. 
\end{mybox}






\section{Design Goals and Overview} \label{section:design1}
Motivated by the unique challenges of agentic inference serving (\S\ref{subsection:challenges}), this section presents the key design goals and a high-level overview of \THISWORK\ toward achieving the optimization objective above.

\subsection{\THISWORK\ Design Goals} \label{subsection:design_goals}
\THISWORK\ defines three \textbf{\underline{D}}esign \textbf{\underline{G}}oals.

\begin{itemize}[leftmargin=*]
    \item \textbf{\underline{DG1:} Context-aware frequency control.} Control GPU frequency using context as a first-class signal so the system can trade speed for power savings when memory headroom exists. The key goal is to \textit{reduce frequency} only up to the point that the \textit{system remains in the non-thrashing regime} while \textit{maintaining SLO attainment}.

    \item \textbf{\underline{DG2:} Context-aware concurrency control.} Regulate how many agents are admitted to each serving instance so that aggregate context demand remains within a safe memory envelope. The goal is to \textit{avoid} overloading any instance with excessive long-lived context that would \textit{trigger thrashing} and hurt both performance and power.

    \item \textbf{\underline{DG3:} Power-aware multi-instance scaling and routing.} For multiple serving instances, scale resources and route requests across them to \textit{jointly minimize system power} and \textit{maintain SLO attainment}. The goal is to consolidate load onto fewer instances under light demand to exploit low idle-power states, while spreading load under higher context pressure to avoid local memory instability.
\end{itemize}

\subsection{\THISWORK\ Design Overview} \label{subsection:design_overview}
\begin{figure}
    \centering
    \includegraphics[width=\linewidth]{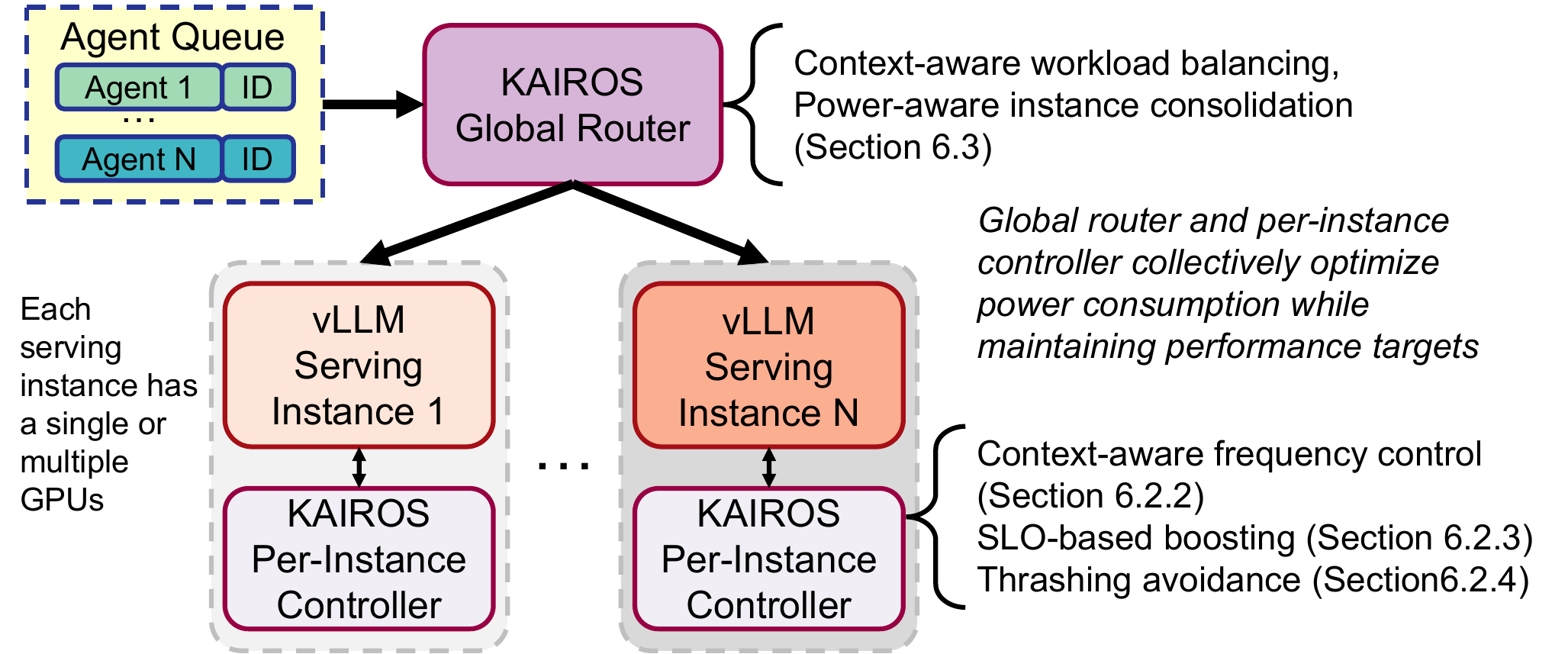}
    \caption{Design overview of \THISWORK: it tracks serving requests for each agent, a global router assigns requests to different vLLM serving instances, and per-instance controller that adjusts GPU frequency to optimize power.}
    \label{fig:design_overview}
\end{figure}
The design goals motivate a unified context-aware runtime that jointly controls frequency, concurrency, and multi-GPU scaling to reduce power while preserving SLO attainment.
Figure~\ref{fig:design_overview} shows the high-level design of \THISWORK.
\THISWORK\ is designed around the three goals in \S\ref{subsection:design_goals} to address the challenges (\S\ref{subsection:challenges}), and has the following components.

\begin{itemize}[leftmargin=*]
    \item \textbf{Agent ID Tracker.}  
    \THISWORK\ first annotates incoming LLM requests with an agent identifier
    so that requests belonging to the same multi turn workflow can be tracked across turns.
    Unlike a semantic variable~\cite{lin2024parrot}, which encode application-level data dependencies and request semantics across LLM calls, our tracking is purely identifier-based and carries no information about request content or relationships.
    This is necessary because, from the viewpoint of the LLM serving system, requests from different agents otherwise appear as ordinary LLM requests without explicit agent-level structure.

    \item \textbf{Context-Aware Per-Instance Controller.}  
    Each serving instance
    is paired with a local controller that monitors context growth, memory pressure, and agent progress, and then manages the instance accordingly. At a high level, this controller adjusts GPU frequency
    and regulates per-instance concurrency to save power, avoid thrashing, and preserve SLO attainment, directly supporting \textit{DG1} and \textit{DG2} while addressing \textit{C1--C3}.
    There is a single controller per serving instance that spans either a single or multiple GPUs, depending on the model size.

    \item \textbf{Context-Aware Multi-instance Router.}  
    Across multiple serving instances, \THISWORK\ employs a global, context-aware router that dynamically adapts instance utilization to workload intensity. 
    Under low load, the router \textit{consolidates} agents onto a subset of instances to keep other GPUs in low-power idle states, while under high load it \textit{spreads} agents to balance context pressure and avoid thrashing. 
    By jointly optimizing placement for both memory stability and power efficiency, the router improves system-wide SLO attainment while reducing unnecessary power overhead, supporting \textit{DG3} to address \textit{C2} and \textit{C3}.

\end{itemize}

Together, these components form a unified runtime that uses context as the central control signal across both local instance management and global request routing.

\section{\THISWORK\ Design Details} \label{section:design2}

In this section, we present the detailed design of \THISWORK.

\subsection{Agent ID Tracker} \label{subsection:design:agent_id_tracker}

A key challenge in agentic LLM inference is that serving backends are not agent-aware: requests are issued independently without agent identifiers, obscuring provenance and preventing agent-level scheduling and control. Modifying agent programs to add such annotations is impractical due to their complexity, heterogeneity, and rapid evolution.

To address this, \THISWORK\ introduces a non-intrusive \emph{Agent ID Generator} that enables agent-level tracking without modifying user programs. A lightweight wrapper launches each agent as a child process, assigns a unique ID, emits lifecycle signals (\texttt{agent\_start}/\texttt{agent\_end}), and transparently encodes the ID into each request (via the API key), ensuring end-to-end propagation. An example interface is shown in Appendix~\ref{app:id-tracker}.

At the backend, the router extracts agent IDs from requests to enable agent-level scheduling, context tracking, and control. This design is fully compatible with existing interfaces, incurs minimal overhead by reusing metadata channels, and provides the abstraction needed for \THISWORK’s context-aware optimizations.

\subsection{Context-Aware Per-Instance Controller} \label{subsection:design:per_instance_controller}
The goal of the \THISWORK\ \emph{per-instance context controller} is to manage each serving instance so that it reduces power while preserving SLO attainment and avoiding the non-ideal thrashing regime.
Its key insight is that, in agentic serving, control decisions such as GPU frequency scaling and concurrency regulation must be driven by \emph{context dynamics}, since each agent’s state persists across turns and directly shapes future memory pressure and performance.
This makes per-instance control fundamentally different from conventional stateless inference, where policies can focus primarily on the latency of the current request.

Accordingly, this subsection presents the controller through five components.
\emph{Control Policy Formulation} defines the local optimization objective.
\emph{Context-Aware Frequency Control} addresses \textit{DG1} and tackles \textit{C1/C3} by trading power for performance based on context pressure.
\emph{SLO-based Frequency Boosting} preserves throughput targets under slowdown, directly addressing \textit{C1}.
\emph{Thrashing Avoidance} provides the safety mechanism that keeps the instance within a stable non-thrashing region and performs context-aware concurrency control (\textit{DG2}).
Together, these components realize the local control plane of \THISWORK, while \textit{DG3} is handled globally by the multi-instance router (\S\ref{subsection:design:context_aware_router}).

\subsubsection{Control Policy Formulation}
\label{sec:control-policy}






We model the controller as a discrete-time policy that operates at control epochs $k = 0, 1, 2, \dots$. 
At each epoch, agents are divided into two sets: the set of ongoing agents $\mathcal{A}_k^{\mathrm{on}}$ that are currently executing, and the set of pending agents $\mathcal{A}_k^{\mathrm{pend}}$ that have arrived but have not yet been admitted.

We define the system context usage at epoch $k$ as the aggregate context footprint of all ongoing agents:
\begin{equation}
U_k = \sum_{a \in \mathcal{A}_k^{\mathrm{on}}} C_{j_a(k)}^a.
\end{equation}

At each control epoch, the controller selects a GPU frequency setting $f_k$ and updates the set of ongoing agents by admitting a subset of $\mathcal{A}_k^{\mathrm{pend}}$ into execution. The control policy is therefore defined as
\begin{equation}
\pi: U_k \mapsto (f_k, \mathcal{A}_k^{\mathrm{on}}).
\end{equation}
Here, $f_k$ governs the power--performance trade-off, while the evolution of $\mathcal{A}_k^{\mathrm{on}}$ regulates concurrency to keep the system within a stable non-thrashing regime.

\subsubsection{Context-Aware Frequency Control}
\label{sec:ctx-aware-policy}

We first describe the frequency control, which addresses \textit{DG1} and directly targets \textit{C1} and \textit{C3} by adapting GPU frequency to the current context pressure of each serving instance.

At instance startup, \THISWORK\ obtains two pieces of static information: 
(i) the set of available GPU frequency levels
\[
\mathcal{F} = \{f^{(1)}, f^{(2)}, \dots, f^{(L)}\},
\]
sorted in ascending order, and 
(ii) the token capacity of the instance, denoted by $U_{\max}$.

At control epoch $k$, the controller observes the aggregate context usage $U_k$ and selects the frequency level as
\begin{equation}
f_k =
\begin{cases}
f^{(L)}, & U_k \ge \alpha U_{\max}, \\[4pt]
f^{(\ell_k)}, & \text{otherwise},
\end{cases}
\end{equation}
where $\alpha \in (0,1]$ is the context-pressure threshold for entering the protective high-frequency regime, and
\begin{equation}
\ell_k = \left\lfloor \frac{U_k}{\alpha U_{\max}} \cdot (L-1) \right\rfloor + 1.
\end{equation}


This policy linearly partitions the safe region $[0, \alpha U_{\max})$ into $L-1$ intervals and increases the frequency level monotonically with context usage.
In practice, we set $\alpha = 0.75$.

The design choice is guided by three principles.
\textbf{First}, when context usage is low, the instance has substantial unused memory capacity.
In this regime, reducing frequency slows individual requests but can safely trade excess capacity for lower power, since the additional context residency still fits within the safe region.
In other words, the controller exploits available context headroom to reduce power.



\textbf{Second}, as context usage rises, the same slowdown becomes increasingly risky.
Lower frequency prolongs agent lifetime, increases resident context, and can accelerate context accumulation, eventually pushing the instance into the thrashing regime.
To address this risk, the controller progressively reduces the aggressiveness of frequency downscaling as $U_k$ grows, using discrete levels to provide a smooth transition from low-pressure to high-pressure operation.

\textbf{Third}, low context usage often coincides with low memory pressure and short per-iteration execution time, which means the instance may be operating well above the performance required to meet the SLO.
In this region, the controller can safely exploit available SLO slack for more aggressive power savings.
Taken together, this policy uses aggregate context as the key runtime signal for balancing power reduction against performance and memory stability.


\subsubsection{SLO-Aware Frequency Boosting}
\label{sec:slo-boost}

While the context-aware frequency policy improves power efficiency by exploiting available context headroom, it does not by itself guarantee per-agent performance.
To address \textit{C1} and further support \textit{DG1}, \THISWORK\ adds a \textit{lightweight runtime safeguard} that boosts frequency when observed agent progress falls below the target SLO.

Specifically, for each agent $a$, we track its achieved throughput up to epoch $k$:
\begin{equation}
\text{Throughput}_k(a) = \frac{\sum_{i \leq j_a(k)} \Delta d_i}{\sum_{i \leq j_a(k)} t_i},
\end{equation}
where $j_a(k)$ denotes the latest completed step of agent $a$ at epoch $k$.


At each control epoch, the controller checks the minimum throughput across all in-process agents:
\begin{equation}
\text{Throughput}_k^{\min} = \min_{a \in \mathcal{A}_k^{\mathrm{on}} \cup \mathcal{A}_k^{\mathrm{pend}}} \text{Throughput}_k(a).
\end{equation}


If $\text{Throughput}_k^{\min}$ falls below the SLO target $\tau$, the controller overrides the context-aware policy and sets the frequency to the maximum level:
\begin{equation}f_k \leftarrow f^{(L)}.\end{equation}
This mechanism serves as a corrective boost when the system begins to under-perform.
It temporarily increases serving capacity to recover throughput.
Under normal conditions, the controller continues to benefit from context-aware power savings.
Under stress, however, it prioritizes performance targets and reduces the likelihood of SLO violations.



\subsubsection{Thrashing Avoidance via Concurrency Control}
\label{sec:thrash-avoidance}

While frequency control helps manage the power--performance trade-off under changing context pressure, it is not sufficient to keep the system out of the context-thrashing regime.
This is challenging because thrashing creates a negative feedback loop: once throughput drops, requests accumulate, resident context grows further, and the resulting memory pressure drives the system even deeper into instability.
Consequently, even a short overload can push the instance into a persistently inefficient operating region.
To address \textit{DG2}, \textit{C2} and \textit{C3}, \THISWORK\ employs a concurrency control mechanism that bounds the aggregate context footprint of ongoing agents.



Specifically, at each control epoch $k$, the controller enforces an upper bound on total context usage:
\begin{equation}
U_k \le \beta U_{\max},
\end{equation}
where $\beta \in (0,1)$ is a safety margin (e.g., $\beta = 0.95$ in practice) that reserves headroom for transient decode growth, which is not directly observable before execution.


If admitting a new agent would violate this constraint, the agent remains in the pending set.
If the constraint is already violated (e.g., due to bursty arrivals), newly arrived agents are deferred to avoid further escalation.
To improve utilization without inducing oscillation, the controller uses a second threshold $\gamma < \beta$ (e.g., $\gamma = 0.9$) for admission.
When pending agents exist and $U_k < \gamma U_{\max}$, the controller incrementally admits agents from $\mathcal{A}_k^{\mathrm{pend}}$ in arrival order into $\mathcal{A}_k^{\mathrm{on}}$ until the threshold is approached.
This two-threshold design keeps concurrency near the safe capacity boundary, preserves headroom for short-term growth, and maintains operation in the non-thrashing regime while still utilizing available memory efficiently.

\subsubsection{Putting It Altogether}
Algorithm~\ref{alg:controller} in \S\ref{appendix:per_instance_pseudo} summarizes the per-instance control loop in \THISWORK. 
At each control epoch, the controller first updates the current context usage, per-agent throughput, and average power, and then selects a baseline frequency using the context-aware policy.
It next applies two corrective mechanisms: SLO-based boosting to recover performance when throughput falls below target, and thrashing avoidance to regulate admissions and keep context usage within a safe memory region.
Together, these steps provide a unified control loop that balances power saving, SLO attainment, and memory stability.

\subsection{\THISWORK\ Context-Aware Multi-Instance Router} \label{subsection:design:context_aware_router}
Modern GPUs exhibit a significant difference in power consumption between idle and active states, rather than a smooth scaling with workload intensity.
In particular, an idle GPU can operate at a very low power state (e.g., 50W on NVIDIA H100), while even a small amount of incoming work can trigger a substantial increase in power (e.g., an additional 100W), even under light load and at the lowest frequency setting.
This creates a significant gap between idle and lightly-loaded operating regimes, making naive policies like round robin less efficient due to activates all instances.

To this end, \THISWORK\ introduces a context-aware router that adapts instance utilization to the input load; its pseudo-code is shown in Algorithm~\ref{alg:router_assign} and \ref{alg:router_reassign} in \S\ref{appendix:global_router}.
Under peak demand, the router tends to utilize all available instances, while under lighter load it consolidates agents onto a smaller subset of GPUs and leaves the remaining instances idle to avoid unnecessary power overhead.\footnote{Even when an instance is idle, it continues running the vLLM server and consumes power close to the GPU’s idle level, which allows it to be brought back into service quickly without incurring a long restart delay~\cite{stojkovic2025dynamollm}.}
The router achieves this through two components: an assignment policy for initial placement and a reassignment policy for dynamic load balancing.

\textbf{\textit{Assignment policy.}}
The router uses a threshold-based assignment policy based on per-instance ongoing agent context usage.
When there exist instances whose ongoing context usage remains below a threshold (set to 50\% of the maximum context capacity), \THISWORK\ prioritizes consolidation by assigning new agents in increasing instance ID order, effectively filling lower-index instances first.
Once all active instances exceed this threshold, the router switches to a load-spreading mode and assigns each incoming agent to the instance with the lowest current context usage.

\textbf{\textit{Reassignment policy.}}
\THISWORK\ maintains, for each agent, a step counter since its last (re)assignment and allows reassignment only after the counter reaches a fixed threshold (8 conversation turns in our implementation).
At that point, the router compares the context usage of the agent's current instance with that of the active instance having the lowest context usage.
If the current instance's context usage is at least 2$\times$ higher, the agent is reassigned to that lower-usage instance, and its counter is reset.
\section{Evaluation Methodology} \label{section:methodology}

We evaluate \THISWORK\ on an NVIDIA H100 NVL server, where each GPU has a 400 W TDP, paired with an Intel Xeon Platinum 8592+ CPU and 2 TB of DRAM.
We use vLLM~\cite{kwon2023pagedattention} v0.14.0 to host the LLM serving instance. Our primary model is \textit{Qwen3-Coder-30B},\footnote{We additionally use \textit{Ministral-3-14B} for model diversity.} quantized to FP8, with each vLLM instance using a single H100 GPU with 80 GB of memory.
We evaluate \THISWORK\ under varying input request arrival rates and SLO targets.
We do not use prefill--decode (PD) disaggregated inference, since context pressure and context-driven serving behavior remain fundamental regardless of disaggregation; \S\ref{appendix:pd_disaggregation} provides a detailed justification.

Our evaluation uses three widely used datasets spanning diverse domains.
(1) \textit{SWE-Bench Verified}~\cite{jimenez2024swebench}, a human validated benchmark for whether agents can generate test-passing fixes for real GitHub issues; (2) \textit{DABStep}~\cite{egg2025dabstep}, a benchmark of real-world data analysis tasks for evaluating multi-step reasoning over structured and unstructured data; and (3) \textit{Terminal-Bench 2.0}~\cite{merrill2026terminalbench}, a benchmark of realistic, human-verified command-line tasks for evaluating aegnt's ability to autonomously complete complex, long-horizon tasks.
We use Zeus~\cite{zeus-nsdi23,chung2025mlenergybenchmarkautomatedinference} for all the frequency and power related control and measurement.


We use the two different agents scaffolding, whose implementation in Harbor~\cite{Harbor_Framework}, for the evaluation:
(1) \textit{mini-swe-agent~\cite{yang2024sweagent}}: a lightweight open-source LLM agent that achieves strong performance despite its minimal design.
(2) \textit{terimus-2~\cite{merrill2026terminalbench}}: Harbor’s autonomous reference agent for evaluating LLM capabilities.

To evaluate the broad applicability of \THISWORK\ across different models and agent harness, we further include experiments based on Ministral-3-14B~\cite{liu2026ministral3} as well as Claude Code~\cite{anthropic2025claudecode} as agent harness, in addition to the main experiments.

For a fair evaluation, we first record the requests and timing from a real agent execution, then replay those requests with exactly the same token distribution, relative timing, and agent issuing order.
As a result, our evaluation guarantees identical workload across different runs, eliminating any runtime variations.
The evaluation runs over a 3-hour window to capture context dynamics and reflect real-world long-running workloads.
\section{Evaluation Results} \label{section:results}
\begin{figure*}
    \centering
    \includegraphics[width=\linewidth]{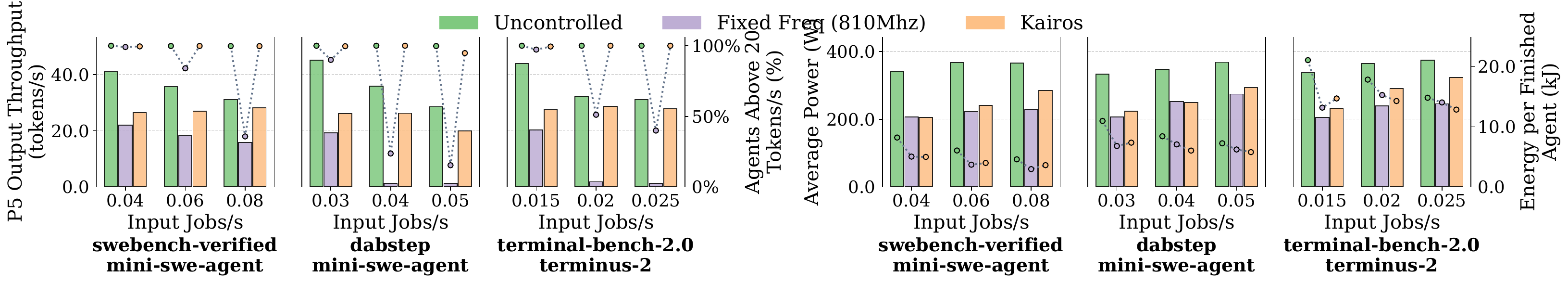}
    \caption{Comparison of throughput, SLO attainment (20 tokens/s per-agent target), energy, and power across request rates for no frequency control, fixed 810 MHz, and \THISWORK. \THISWORK\ reduces power by 27\% on average without sacrificing SLO.}
    \label{fig:performance_power_main_result}
\end{figure*}
This section presents a detailed evaluation of how \THISWORK\ design choices impact both performance and power.

\subsection{Single-Instance Agentic Serving}
\textbf{Performance and power comparison.}
Figure~\ref{fig:performance_power_main_result} compares performance, power, and energy across different agent jobs/s arrival rates, with a per-agent SLO target of 20 tokens/s.
We present additional results under varying SLO targets in the following discussion.
In this section, we use P5 throughput as the performance metric, defined as the throughput of the slowest 5\% of agents; analogous to P95 latency.
The study evaluates three baselines: no frequency control, a fixed GPU frequency of 810 MHz, and \THISWORK.
While both fixed frequency baseline and \THISWORK\ reduce power relative to no control, the fixed-frequency baseline achieves slightly larger power savings at the cost of significantly degraded SLO attainment.
In contrast, \THISWORK\ strikes a balance by achieving an average power saving of 27\% (up to 39.8\%) while meeting the per-agent SLO target of 20 tokens/second.

This result \textit{underscores the effectiveness of a dynamic solution} over a static approach.
While statically reducing GPU frequency can lower power, it is insufficient on its own because it often causes significant SLO violations.
Even from a power-only standpoint, it is impractical for deployment since it requires expensive per-workload grid search and frequency tuning for finding an optimal operating conditions, which keeps changing rapidly under a highly dynamic workload  (\S\ref{section:characterization}).
Instead, \THISWORK\ achieves a careful balance between power reduction and performance, maintaining SLO attainment while reducing power.
This balance comes from dynamically adapting GPU frequency and concurrency based on context pressure and agent progress through context-aware frequency control, SLO-driven boosting, and thrashing-aware concurrency control.

\noindent
\textbf{Effectiveness of SLO boosting with different targets.}
\begin{figure}
    \centering
    \includegraphics[width=\linewidth]{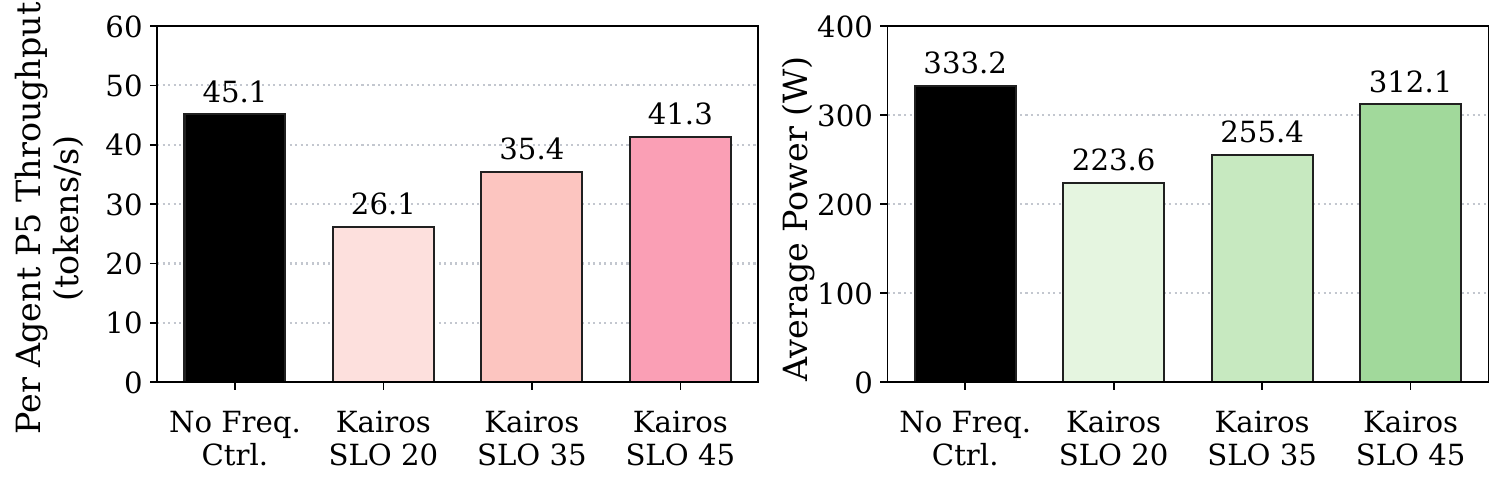}
    \caption{\textbf{Effectiveness of SLO boosting.} P5 throughput and average power of \THISWORK\ with different SLO targets of 20, 35, and 45 tokens/s. This experiment runs mini-swe-agent on DABStep with an arrival rate of 0.03 jobs/s.}
    \label{fig:different_slo}
\end{figure}
Figure~\ref{fig:different_slo} shows per-agent P5 throughput (left) and average power (right) under different SLO targets: 20, 35, and 45 tokens/s at a fixed request arrival rate of 0.03 agent jobs/s.
This data is collected by running mini-swe-bench with DABstep.
It compares a no frequency  control baseline with \THISWORK\ configured for different SLO targets.
Three key observations emerge. 
\textit{First}, \THISWORK\ adapts effectively to different SLO targets, meeting the required throughput while significantly reducing power.
This demonstrates the effectiveness of its context-aware frequency control and SLO-driven boosting mechanisms of \THISWORK.

\textit{Second}, tighter SLO targets require higher GPU frequencies, which reduces power savings: as the SLO increases from 20 to 45 tokens/s, power savings drop from 32.9\% to 6.3\% relative to the baseline.
The 45 tokens/s target is close to the full-frequency hardware limit. At this aggressive SLO, \THISWORK\ barely meets the target, leaving limited opportunity for frequency reduction and therefore yielding a less favorable power--performance trade-off.
Appendix \S\ref{appendix:agent_distribution} further shows per-agent throughput distribution.

\textit{Third}, the figure reveals a hardware limit, where 45 tokens/s represents the maximum achievable per-agent SLO for this setup.
Pushing beyond this point would significantly degrade SLO attainment.
This indicates a tipping point governed by hardware capacity, suggesting that further improvements in SLO would require additional compute resources rather than better control alone.

\begin{figure}
    \centering
    \includegraphics[width=\linewidth]{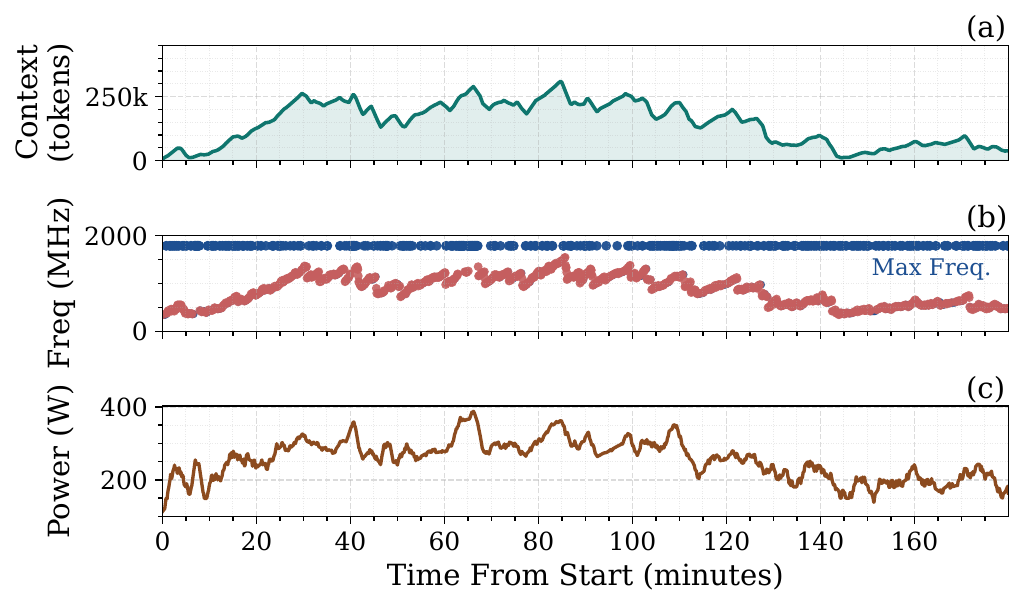}
    \caption{Change in instantaneous power, GPU frequency, and context size of \THISWORK\ with respect to time for mini-swe-agent running on DABStep with arrival rate of 0.03 jobs/s and SLO target of 35 tokens/s.}
    \label{fig:power_freq_context_time}
\end{figure}
\begin{figure}
    \centering
    \includegraphics[width=\linewidth]{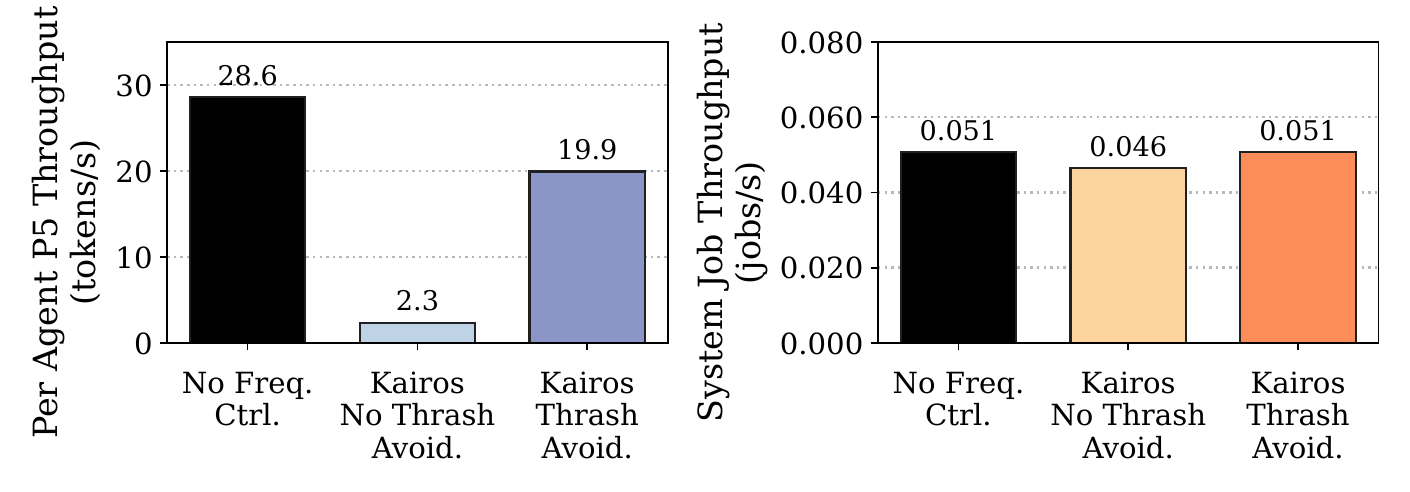}
    \caption{\textbf{Effectiveness of thrashing avoidance}. P5 throughput and overall job throughput for mini-swe-agent on DABStep with request rate of 0.05 jobs/s and an SLO target of 20 tokens/s.}
    \label{fig:thrashing_avoidance}
\end{figure}
\begin{figure}
    \centering
    \includegraphics[width=\linewidth]{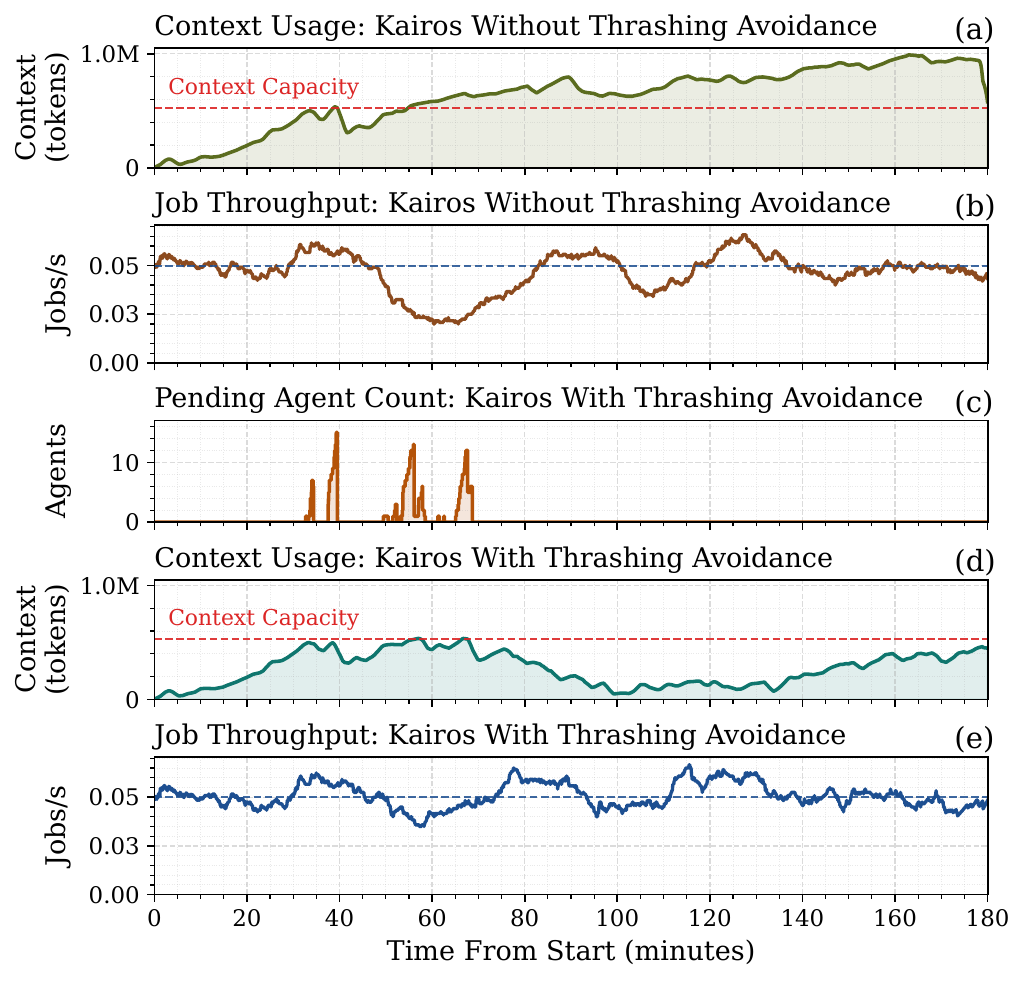}
    \caption{Context size, agent processing throughput and pending agents with and without employing thrashing avoidance technique in \THISWORK. This experiment runs mini-swe-agent with DABStep with a request rate of 0.05 jobs/s and a target SLO of 20 tokens/s.}
    \label{fig:thrashing_avoidance_detail}
\end{figure}

Figure~\ref{fig:power_freq_context_time} provides a deeper look into \THISWORK’s runtime behavior under a request arrival rate of 0.03 jobs/s\footnote{Due to space limitation, we carefully select representative request rates and SLO targets to demonstrate the effectiveness of different techniques proposed in \THISWORK\ in Figures~\ref{fig:power_freq_context_time}, \ref{fig:thrashing_avoidance}, and \ref{fig:thrashing_avoidance_detail}.} and an SLO target of 35 tokens/s by showing instantaneous power, GPU frequency, and context size over time.
As context size (subfigure (a)) grows, \THISWORK\ gradually increases GPU frequency (subfigure (b)) to accelerate agent progress and prevent excessive accumulation, while reducing frequency when context pressure subsides to save power (subfigure (c)).
This adaptive behavior is reflected in the corresponding changes in power consumption, which closely track frequency adjustments.
When the system detects potential SLO violations, it temporarily boosts the GPU to maximum frequency (shown by blue points) to recover performance.
Overall, the figure illustrates how \THISWORK\ dynamically responds to changing workload conditions to balance power savings and SLO attainment using context as a first-class control signal.

\noindent
\textbf{Effectiveness of thrashing avoidance (concurrency control).}
Figure~\ref{fig:thrashing_avoidance} evaluates the effectiveness of the thrashing avoidance mechanism under a higher load (0.05 jobs/s) with an SLO target of 20 tokens/s using mini-swe-agent on DABStep.
It compares three configurations: no frequency control, \THISWORK\ without thrashing avoidance, and \THISWORK\ with thrashing avoidance.
While the no-control baseline avoids thrashing by running at high frequency, \THISWORK\ without thrashing avoidance reduces frequency to save power, pushing the system into a thrashing regime.
This leads to a severe drop in per-agent P5 throughput (left figure) to 2.3 tokens/s and overall throughput (right figure) of 0.046 jobs/s, failing to keep up with the arrival rate.
In contrast, enabling thrashing avoidance mechanism allows \THISWORK\ to maintain P5 throughput close to the 20 tokens/s target and sustain system throughput at 0.05 jobs/s, demonstrating stable and efficient operation.

Figure~\ref{fig:thrashing_avoidance_detail} provides a time-series view explaining this behavior.
Without thrashing avoidance, context usage (subfigure (a)) exceeds capacity and remains high, pushing the system into a persistent thrashing regime where throughput drops (subfigure (b)) and queued agents accumulate (subfigure (c)), making recovery difficult.
This leads to unstable operation as the system cannot drain incoming work fast enough.
With thrashing avoidance enabled, \THISWORK\ prevents this buildup by controlling context growth (subfigure (d)) and, when necessary, triggering corrective frequency boosts to reduce accumulated context.
As a result, context stays within capacity, job throughput remains stable (subfigure (e)), and avoids queue buildup.
These results highlight how thrashing avoidance and SLO-aware boosting work together to maintain system stability under high load.

\noindent
\textbf{Effectiveness across multiple models.}
\begin{table}[t]
\centering
\scriptsize
\caption{Effectiveness of \THISWORK\ across multiple models.}
\label{tab:power}
\begin{tabular}{lcc|cc}
\toprule
 & \multicolumn{2}{c}{Qwen3-Coder-30B (0.06 jobs/s)} & \multicolumn{2}{c}{Ministral-3-14B (0.06 jobs/s)} \\
\cmidrule(lr){2-3} \cmidrule(lr){4-5}
 & No Freq. Control & \THISWORK & No Freq. Control & \THISWORK \\
\midrule
Power (W) & 367.9 & 240.9 & 370.8 & 263.8 \\
\bottomrule
\end{tabular}
\end{table}
Table~\ref{tab:power} shows that the benefits of \THISWORK\ generalize across models, both running SWE-Bench Verified.
On \textit{Qwen3-Coder-30B}, \THISWORK\ reduces power by 34.5\%, and on \textit{Ministral-3-14B}, it reduces power by 28.9\%.
These results indicate that the benefit of \THISWORK\ is not tied to a single model, but is broadly effective across different LLMs.


\subsection{Generality Across Agent Harnesses and Models}

We further broaden the evaluation using a production agent harness, Claude Code, and another model, Ministral-3-14B~\cite{liu2026ministral3}.
Table~\ref{tab:claude_code} evaluates \THISWORK\ with Claude Code under different request loads.
Across the evaluated loads, \THISWORK\ reduces average GPU power by 9.0--36.7\%, showing that its benefits extend to a production agent harness beyond simple ReAct-style agents.
Table~\ref{tab:ministral} evaluates \THISWORK\ with Ministral-3-14B on DABStep and Terminal-Bench 2.0.
\THISWORK\ reduces average GPU power by 26.1\% on DABStep and 21.9\% on Terminal-Bench 2.0, demonstrating that its effectiveness is not tied to a specific model.
Together, these results show that using context usage signal for reducing power generalizes beyond simple ReAct. 
Because \THISWORK\ controls relative to available KV-cache capacity, it is also largely decoupled from a specific model or hardware platform.

\begin{table}[t]
\centering
\scriptsize
\setlength{\tabcolsep}{4pt}
\renewcommand{\arraystretch}{1.05}
\caption{Power reduction with Claude Code under different request loads (JPS: jobs per second).}
\label{tab:claude_code}
\begin{tabular}{lrrr}
\toprule
Load & Baseline & \THISWORK & Power Reduction \\
\midrule
0.001 JPS & 354.71 W & 224.45 W & 36.72\% \\
0.002 JPS & 385.74 W & 292.13 W & 24.27\% \\
0.005 JPS & 390.12 W & 354.94 W & 9.02\% \\
\bottomrule
\end{tabular}
\end{table}

\begin{table}[t]
\centering
\scriptsize
\setlength{\tabcolsep}{4pt}
\renewcommand{\arraystretch}{1.05}
\caption{Power reduction with Ministral-3-14B across additional benchmarks (JPS: jobs per second).}
\label{tab:ministral}
\begin{tabular}{lrrr}
\toprule
Benchmark / Load & Baseline & \THISWORK & Power Reduction \\
\midrule
DABStep, 0.03 JPS
    & 341.57 W & 252.46 W & 26.09\% \\
Terminal-Bench 2.0, 0.015 JPS
    & 334.56 W & 261.23 W & 21.92\% \\
\bottomrule
\end{tabular}
\end{table}

\subsection{Multi-Instance Agentic Serving}
\begin{figure}
    \centering
    \includegraphics[width=\linewidth]{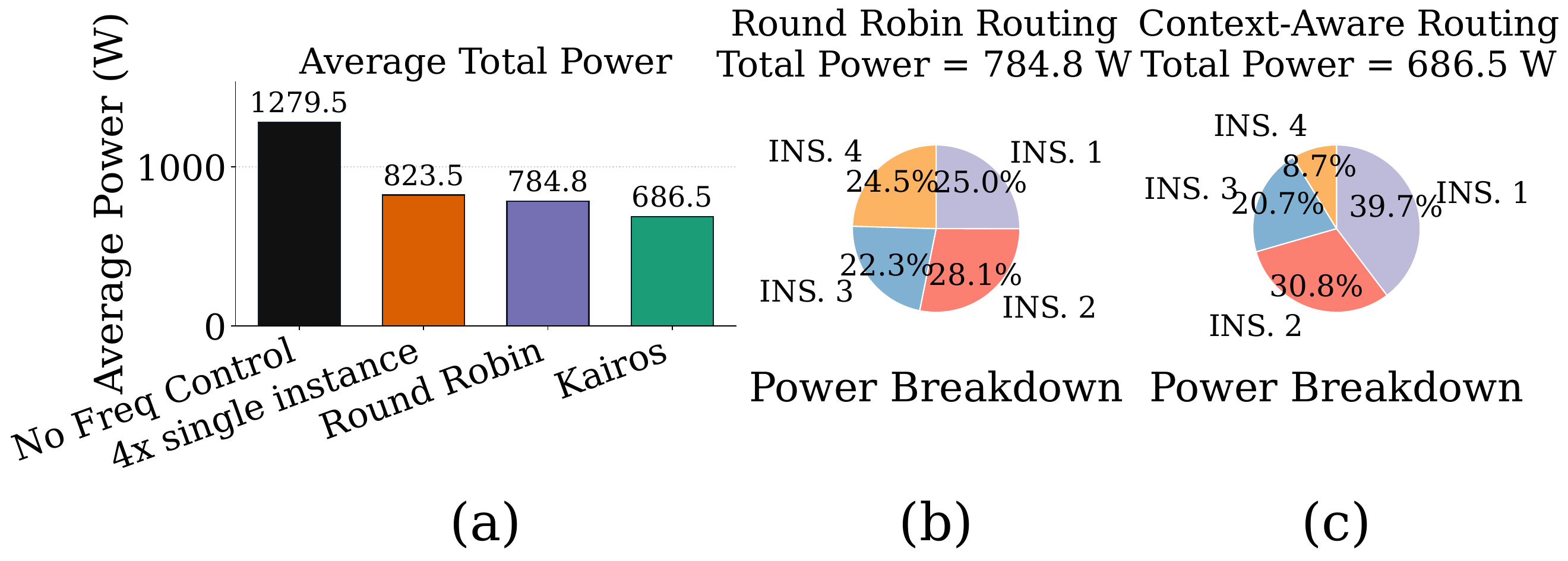}
    \caption{(a) Power comparison across four serving instances for no frequency control, a replicated single-instance baseline, a round-robin routing policy, and \THISWORK\ with context-aware routing. (b,c) Per-instance power breakdown for round robin and \THISWORK\ routing.}
    \label{fig:multi_instance_serving}
\end{figure}
Figure~\ref{fig:multi_instance_serving} demonstrates the effectiveness of \THISWORK\ under multi-instance scaling at an input rate of 0.16 jobs/s.
This experiment runs mini-swe-agent using SWE-Bench Verified dataset.
Compared to the baseline without frequency control, \THISWORK\ with round-robin routing reduces average power from 1279.5~W to 784.8~W, while its context-aware routing further lowers it to 686.5~W, achieving a total power reduction of 46.3\%.
The key difference lies in how work is distributed across instances: round-robin assigns load evenly, preventing GPUs from entering low-power idle states, whereas \THISWORK’s context-aware routing consolidates work onto fewer GPUs when possible.
This consolidation avoids the significant power overhead of lightly-loaded GPUs and enables other instances to remain near idle power. 
As shown in Figures~\ref{fig:multi_instance_serving}(b,c), this results in a more efficient power distribution across instances, highlighting the benefit of coordinated routing and control.
In addition, \THISWORK's context-aware router achieves per agent p5 throughput of 26.9 tokens/s, surpassing the 25.4 tokens/s when using round robin policy.

\subsection{Runtime Overhead}

\THISWORK\ only tracks lightweight per-agent metadata and routes agents across instances, so the added work is limited to bookkeeping and routing operations.
Therefore, the tracking, placement, and routing overhead introduced by \THISWORK\ is small. 
Table~\ref{tab:kairos_overhead} reports the measured overhead across the three benchmarks.
The overhead remains around 3\% in all cases, confirming that \THISWORK\ introduces only modest runtime cost.

\begin{table}[t]
\centering
\scriptsize
\setlength{\tabcolsep}{5pt}
\renewcommand{\arraystretch}{1.05}
\caption{Runtime overhead of \THISWORK.}
\label{tab:kairos_overhead}
\begin{tabular}{lcc}
\toprule
Benchmark & Overhead & Percentage \\
\midrule
DABStep & 0.164 s & 3.28\% \\
SWE-bench Verified & 0.142 s & 3.29\% \\
Terminal-Bench 2.0 & 0.180 s & 2.90\% \\
\bottomrule
\end{tabular}
\end{table}

\subsection{Parameter Sensitivity}

\begin{figure}
    \centering
    \includegraphics[width=\linewidth]{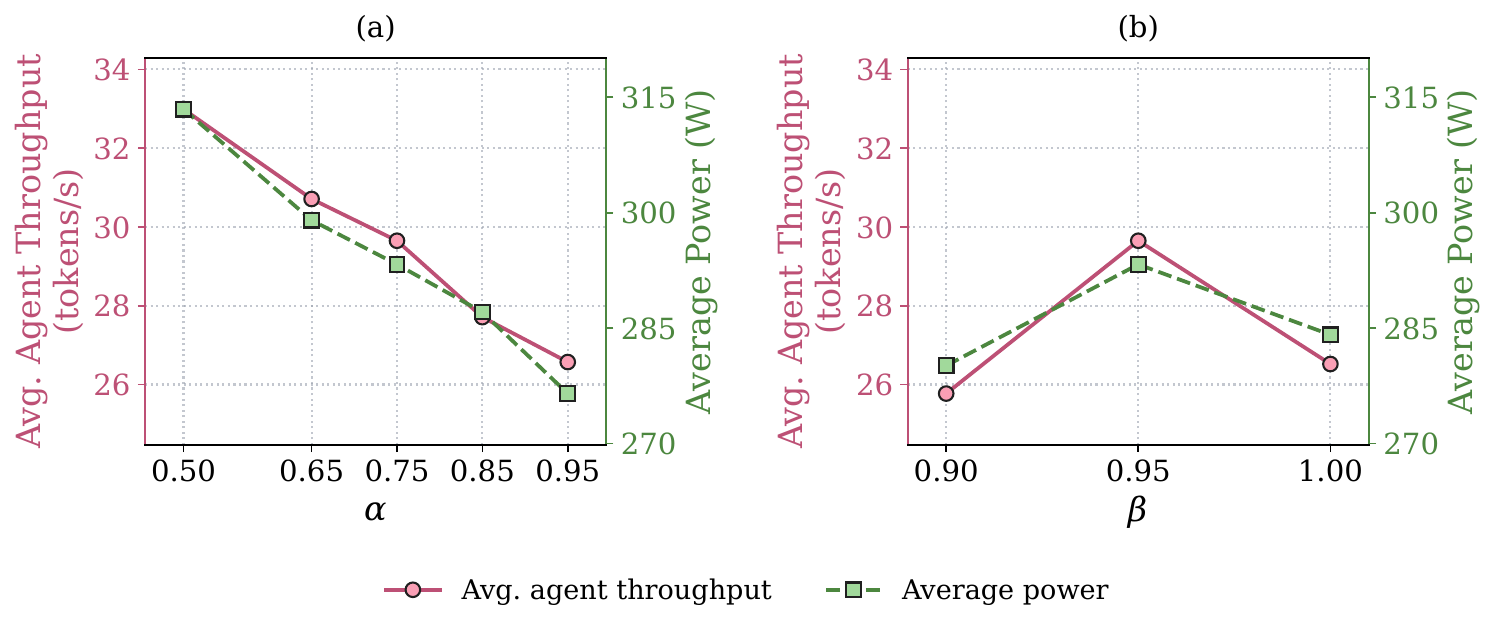}
    \caption{Impact of control parameters $\alpha$ and $\beta$ in \S~\ref{sec:control-policy} on agent throughput and power consumption. (a) $\alpha$: frequency-scaling threshold, balancing power savings and throughput. (b) $\beta$: admission threshold, balancing utilization and thrashing avoidance. }
    \label{fig:ab-param}
\end{figure}



We evaluate the sensitivity of \THISWORK\ to its two control policy parameters in \S\ref{sec:control-policy}, $\alpha$ and $\beta$, which govern frequency scaling and KV-cache headroom, respectively.
The parameter $\alpha$ controls the power--performance trade-off: a higher $\alpha$ saves more power but can trigger more admission control and reduce throughput, while a lower $\alpha$ keeps the system at higher frequency and saves less power. Figure~\ref{fig:ab-param} (a) reports the sensitivity to $\alpha$.
As $\alpha$ increases, \THISWORK\ achieves lower average power at the cost of lower agent throughput. We use $\alpha=0.75$ as a balanced setting between power saving and performance.
The parameter $\beta$ controls how conservatively \THISWORK\ reserves KV-cache headroom. A value that is too high can lead to oscillation as agent states grow and amplify overhead, while a value that is too low underutilizes the serving instance. Figure~\ref{fig:ab-param} (b) reports the sensitivity to $\beta$.
Among the evaluated settings, $\beta=0.95$ achieves the highest average agent throughput while maintaining moderate power consumption. We therefore use $\beta=0.95$ as the default setting.

\subsection{\THISWORK's Impacts on Task Failure Rate}

\begin{table}[t]
\centering
\scriptsize
\setlength{\tabcolsep}{3pt}
\renewcommand{\arraystretch}{1.05}
\caption{Timeout rate and average power on SWE-bench Verified.}
\label{tab:timeout}
\begin{tabular}{crrrr}
\toprule
& \multicolumn{2}{c}{No SLO} &
  \multicolumn{2}{c}{SLO = 25 tokens/s} \\
\cmidrule(lr){2-3}\cmidrule(lr){4-5}
Jobs/s & Timeout & Power & Timeout & Power \\
\midrule
0.04 & 0.00\% (0/454) & 206.64 W &
       0.00\% (0/454) & 235.33 W \\
0.08 & 0.12\% (1/861) & 287.51 W &
       0.00\% (0/861) & 310.32 W \\
\bottomrule
\end{tabular}
\end{table}

Because \THISWORK\ only adjusts execution frequency without modifying model outputs, its impact on task success can arise only indirectly through execution-time constraints such as task timeouts. 
We therefore evaluate the excessive tasks failure on SWE-bench Verified introduced by \THISWORK. 
As shown in Table~\ref{tab:timeout}, at arrival rates of 0.04 and 0.08 jobs/s, \THISWORK\ without an SLO incurs additional task failure rates due to timeout of 0.00\% (0/454) and 0.12\% (1/861), respectively, whereas enforcing a 25 tokens/s SLO results in no additional failures in either setting. 
These results indicate that the throughput-based SLO effectively prevents timeout-induced degradation in task completion.

\section{Discussion}


We discuss several aspects of \THISWORK\ that clarify its scope and relationship to existing serving techniques, including its interaction with KV-cache offloading, the use of throughput as an online SLO for agent progress, and its complementarity with fine-grained DVFS.

\paragraph{Relationship to KV-cache offloading}
\THISWORK's context-aware DVFS is compatible with systems with KV-cache offloading enabled. 
While offloading allows the system to handle higher request loads, our LMCache measurements in Appendix~\S A.4 show that it still introduces noticeable overhead, which ongoing/recent works such as MORI~\cite{xia2026idlenessrelativeexploitingtoolcall}, Continuum~\cite{li2025continuum} and more~\cite{tiwari2026cachewiseunderstandingworkloadsoptimizing, kang2026thunderagent} seek to reduce. 
Therefore, DVFS with offloading enabled still needs to be context-aware: lowering GPU frequency can prolong agent lifetimes, increase aggregate context pressure, and induce additional offload/reload overhead that may offset the benefits from power saving. 
Joint optimization of DVFS and KV-cache offloading is an interesting direction for future work.

\paragraph{The choice of throughput-based SLO}
Agentic workloads are goal-driven: users specify a high-level task, the agent autonomously explores toward completion, and the user-visible latency is end-to-end task-completion time. 
\THISWORK\ uses per-agent generation throughput as an online proxy for end-to-end latency and sustained task progress, since higher token/s reduces end-to-end execution time while task-completion latency is unavailable at runtime.
Fine-grained token-level latency such as per-turn TTFT/TBT is less visible to users during agentic execution. 

\paragraph{Relationship to fine-grained DVFS}
\THISWORK\ complements fine-grained DVFS policies rather than replacing them.
Existing approaches~\cite{stojkovic2025dynamollm,kakolyris2024throttllem,greenllm_dac,basit2026biscale,yu2025voltanallmfeedbackdrivenfrequencycontrol,wang2025usinganalytical,qiu2024muserve,spaan2026kerneldvfs} use local metrics such as TPOT or operator latency, while \THISWORK\ provides a longer-term context-pressure signal that captures their impact on KV-cache residency.

\section{Related Work} \label{section:related_work}
To the best of our knowledge, \textbf{\textit{\THISWORK\ is the first work to optimize power for agentic inference serving}}.
Below, we compare \THISWORK\ with closely related works.

\paragraph{Power optimization in LLM serving.}
Prior work on power-efficient LLM serving mainly falls into two categories: DVFS~\cite{stojkovic2025dynamollm,kakolyris2024throttllem,greenllm_dac,basit2026biscale,yu2025voltanallmfeedbackdrivenfrequencycontrol,wang2025usinganalytical,qiu2024muserve,spaan2026kerneldvfs}, which adjusts GPU frequency online using runtime signals such as slack, batch behavior, KV-cache growth, or phase-specific latency predictions, and system-level scheduling~\cite{wilkins2024hybrid,li2025ecoserve,stojkovic2025tapas}, which reduces power through hardware-aware placement and resource allocation.
These works target mostly \textit{stateless} LLM serving, where requests are treated as independent.
In contrast, agentic serving is fundamentally \textbf{\textit{stateful}}: multi-turn conversations continuously grow and revisit context, so power optimization must account for the evolution of that state over time, which is the key focus of our work.
How to generalize and integrate the idea of \THISWORK\ to energy-first system designs as visioned in~\cite{chowdhury2026energycalculuscompositionalalgebra} is an interesting future work.

\paragraph{Power optimization in LLM training.}
Parallel to inference-side power optimization, another line of work focuses on reducing the power and energy consumption of LLM training~\cite{Chung_2024,wu2026kareusjointreductiondynamic,spaan2026kerneldvfs}. Although some of the underlying mechanisms overlap with \THISWORK, these techniques are tailored to the forward and backward passes of training and do not account for the evolving state inherent in agentic serving like \THISWORK.

\paragraph{Energy/power characterization.}
Several recent works~\cite{stojkovic2024towardsgreenerllms,wilkins2024offline,argerich2024measuring,maliakel2025characterizing,husom2024price,ozcan2025quantifying,niu2026tokenpowerbench,wilhelm2025beyond} characterize, model, or benchmark the energy and power behavior of LLM inference rather than directly optimizing it.
They study inference energy trade-offs and control knobs, build workload-based energy/runtime models, profile fine-grained power use, quantify energy and carbon emissions, and propose benchmarking or reporting frameworks such as energy-per-token.
A closely related work~\cite{kim2026cost} on agents characterizes the infrastructure cost of dynamic reasoning and shows that agentic workflows introduce substantially higher resource and energy demands than conventional single-shot inference.
In contrast, our work takes a step beyond characterization: we design a runtime mechanism that uses context growth as a control signal to actively reduce energy while maintaining safe, non-thrashing performance for agentic serving.

\paragraph{Agentic inference and stateful LLM serving.}
Recent systems optimize agentic workflows through workflow-aware runtimes and program semantics~\cite{lin2024parrot,santhanam2024alto,chaudhry2025murakkab,ro2025sherlock,kang2026thunderagent,raj2025cpu}, while others address multi-round behavior through cache retention and adaptive prefill/decode placement~\cite{li2025continuum,he2026ampd}. In parallel, LLM serving work improves memory efficiency through KV-cache management, paging, offloading, scheduling, and disaggregation~\cite{kwon2023pagedattention,prabhu2025vattention,patel2024splitwise,zhong2024distserve,agrawal2024medha,kim2025infersave}.
Together, these works show stateful and memory-bound workload behavior, but focus mainly on performance. 
In contrast, our work targets \textit{power optimization} in agentic workloads and explicitly accounts for the feedback loop between slower execution, increased agent lifetime, and growing context footprint, using this coupling to drive global control decisions.

\paragraph{Autoscaling and resource management for AI serving.}
Prior work on autoscaling and datacenter resource management provisions resources using signals such as queue length, utilization, token rate, and SLO violations~\cite{patke2025chiron,jaiswal2025sageserve,lai2025tokenscale,singh2025elasticmoe,crankshaw2020inferline,delimitrou2014quasar,lo2015heracles,hadary2020protean}. These methods match capacity to demand, but do not account for the stateful, highly dynamic nature of agentic workloads. In our setting, scaling must also control multi-turn context growth and keep per-instance demand within a safe non-thrashing regime.

\section{Conclusion} \label{section:conclusion}
This paper showed that agentic AI serving is fundamentally different from single-turn LLM serving due to its long-lived, dynamically evolving context and the emergence of a thrashing regime under power scaling.
We found that reducing GPU frequency can unexpectedly worsen both performance and power efficiency by increasing context residency and triggering memory instability.
Based on these insights, this paper designed \THISWORK, a dynamic system that jointly manages frequency, concurrency, and routing using \textit{context as a first-class signal}.
Our results demonstrated that \THISWORK\ enables significant average power savings of 27\% while maintaining performance targets, highlighting the need to rethink power reduction for stateful agentic AI serving.
This work opens the door to a broader class of context-aware, power-efficient runtimes for agentic AI.
It further suggests that rethinking serving around state, memory stability, and dynamic control will be essential as agentic workloads continue to scale.

\begin{acks}
We used AI tools for code development, figure plotting, and text editing. We thank the anonymous reviewers for their constructive feedback and valuable suggestions. This work was supported in part by NSF grants CCF-2450085 and CNS-2106184 and DARPA ML2P Award HR0011-26-9-E190. This work was also supported in part by the Semiconductor Research Corporation (SRC) under the GRC AIHW grant, the AMD AI \& HPC Cluster Program, and Advanced Micro Devices, Inc. through the Funding Academic Research program.
\end{acks}

\bibliographystyle{ACM-Reference-Format}
\balance
\bibliography{01_agentb-bib}

\clearpage

\balance

\appendix

\section{Appendix}

\subsection{\THISWORK\ Per-Instance Controller Pseudo-code} \label{appendix:per_instance_pseudo}
\begin{algorithm}[t]
\caption{\THISWORK\ Per-Instance Controller Algorithm}
\label{alg:controller}
\begin{algorithmic}[1]
\REQUIRE Frequency levels $\mathcal{F} = \{f^{(1)}, \dots, f^{(L)}\}$; capacity $U_{\max}$; 
SLO target $\tau$; power budget $P_{\max}$; thresholds $\alpha, \beta, \gamma$
\ENSURE Frequency level $\ell$ and active set $\mathcal{A}^{\mathrm{on}}$ at each epoch

\STATE Initialize $\ell \gets L$, $c \gets 0$, $\mathcal{A}^{\mathrm{on}} \gets \emptyset$, $\mathcal{A}^{\mathrm{pend}} \gets \emptyset$

\LOOP
    \STATE $\mathcal{A}^{\mathrm{pend}} \gets \mathcal{A}^{\mathrm{pend}} \cup \text{new tasks}$

    \STATE $U_k \gets \sum_{a \in \mathcal{A}^{\mathrm{on}}} C^a$
    \STATE Update per-agent throughput for all $a \in \mathcal{A}^{\mathrm{on}} \cup \mathcal{A}^{\mathrm{pend}}$
    \STATE Update average system power $\bar{P}$

    \STATEX \textcolor{blue}{\textit{// Context-aware frequency control (\S\ref{sec:ctx-aware-policy})}}
    \IF{$U_k \ge \alpha U_{\max}$}
        \STATE $\ell \gets L$
    \ELSE
        \STATE $\ell \gets \left\lfloor \frac{U_k}{\alpha U_{\max}} \cdot (L-1) \right\rfloor + 1$
    \ENDIF

    \STATEX \textcolor{blue}{\textit{// SLO-based boosting (\S\ref{sec:slo-boost})}}
    \STATE $\tau_{\min} \gets \min_{a \in \mathcal{A}^{\mathrm{on}} \cup \mathcal{A}^{\mathrm{pend}}} \text{Throughput}(a)$
    \IF{$\tau_{\min} < \tau$}
        \STATE $\ell \gets L$
    \ENDIF

    \STATEX \textcolor{blue}{\textit{// Thrashing avoidance (\S\ref{sec:thrash-avoidance})}}
    \WHILE{$\mathcal{A}^{\mathrm{pend}} \neq \emptyset$ \textbf{and} $U_k < \gamma U_{\max}$}
        \STATE Move next agent from $\mathcal{A}^{\mathrm{pend}}$ to $\mathcal{A}^{\mathrm{on}}$
        \STATE Update $U_k$
    \ENDWHILE

    \IF{$U_k > \beta U_{\max}$}
        \STATE Defer new admissions
    \ENDIF


    \STATE Apply frequency level $\ell$
\ENDLOOP
\end{algorithmic}
\label{algo:ctrl}
\end{algorithm}
Algorithm~\ref{algo:ctrl} presents the pseudo-code for algorithm that governs per-instance frequency controller in \THISWORK\ that employs context-aware frequency control, SLO-based boosting, and thrashing avoidance.

\subsection{\THISWORK\ Global Router Pseudo-code} \label{appendix:global_router}

\begin{algorithm}[t]
\caption{\THISWORK\ Initial Assignment Algorithm}
\label{alg:router_assign}
\begin{algorithmic}[1]
\REQUIRE Serving instances $\mathcal{I} = \{1,\dots,N\}$; per-instance capacity $U_{\max}$; consolidation threshold $\theta_{\mathrm{cons}}$
\ENSURE Assigned instance $\mathrm{inst}(a)$ for each newly arrived agent $a$

\STATE Initialize $\mathcal{A}_i \gets \emptyset$ for all $i \in \mathcal{I}$

\FUNCTION{AssignAgent}{$a$}
    \STATE Update ongoing context usage $U_i \gets \sum_{b \in \mathcal{A}_i} C^b$ for all $i \in \mathcal{I}$

    \STATEX \textcolor{blue}{\textit{// Initial assignment policy (\S\ref{subsection:design:context_aware_router})}}
    \STATE $\mathcal{I}_{\mathrm{light}} \gets \{i \in \mathcal{I} \mid U_i < \theta_{\mathrm{cons}} U_{\max}\}$
    \IF{$\mathcal{I}_{\mathrm{light}} \neq \emptyset$}
        \STATE $i^\star \gets \min \mathcal{I}_{\mathrm{light}}$
    \ELSE
        \STATE $i^\star \gets \arg\min_{i \in \mathcal{I}} U_i$
    \ENDIF

    \STATE Assign agent $a$ to instance $i^\star$
    \STATE $\mathcal{A}_{i^\star} \gets \mathcal{A}_{i^\star} \cup \{a\}$
    \STATE $\mathrm{inst}(a) \gets i^\star$
    \STATE $s_a \gets 0$
\ENDFUNCTION
\end{algorithmic}
\end{algorithm}

\begin{algorithm}[t]
\caption{\THISWORK\ Reassignment Algorithm}
\label{alg:router_reassign}
\begin{algorithmic}[1]
\REQUIRE Serving instances $\mathcal{I} = \{1,\dots,N\}$; reassignment interval $T_{\mathrm{reassign}}$; imbalance threshold $\rho$
\ENSURE Updated assignment $\mathrm{inst}(a)$ when agent $a$ issues a request

\FUNCTION{MaybeReassign}{$a$}
    \STATE $s_a \gets s_a + 1$

    \IF{$s_a < T_{\mathrm{reassign}}$}
        \STATE \RETURN
    \ENDIF

    \STATE Update ongoing context usage $U_i \gets \sum_{b \in \mathcal{A}_i} C^b$ for all $i \in \mathcal{I}$

    \STATEX \textcolor{blue}{\textit{// Reassignment policy (\S\ref{subsection:design:context_aware_router})}}
    \STATE $i \gets \mathrm{inst}(a)$
    \STATE $j \gets \arg\min_{m \in \mathcal{I}} U_m$

    \IF{$j \neq i$ \textbf{and} $U_i \geq \rho \cdot U_j$}
        \STATE Reassign agent $a$ from instance $i$ to instance $j$
        \STATE $\mathcal{A}_i \gets \mathcal{A}_i \setminus \{a\}$
        \STATE $\mathcal{A}_j \gets \mathcal{A}_j \cup \{a\}$
        \STATE $\mathrm{inst}(a) \gets j$
    \ENDIF

    \STATE $s_a \gets 0$
\ENDFUNCTION
\end{algorithmic}
\end{algorithm}

Algorithm~\ref{alg:router_assign} and Algorithm~\ref{alg:router_reassign} together present the pseudo-code for \THISWORK’s global context-aware router, which performs initial agent placement and dynamic reassignment across serving instances to enable consolidation under low load and balance context pressure under high load.
The detailed explanation for the Algorithm~\ref{alg:router_assign} and \ref{alg:router_reassign} is presented in details in \S\ref{subsection:design:context_aware_router}.

\newpage

\subsection{Context Usage Variation} \label{appendix:context_variation}
\begin{figure}[t]
    \centering
    \includegraphics[width=\linewidth]{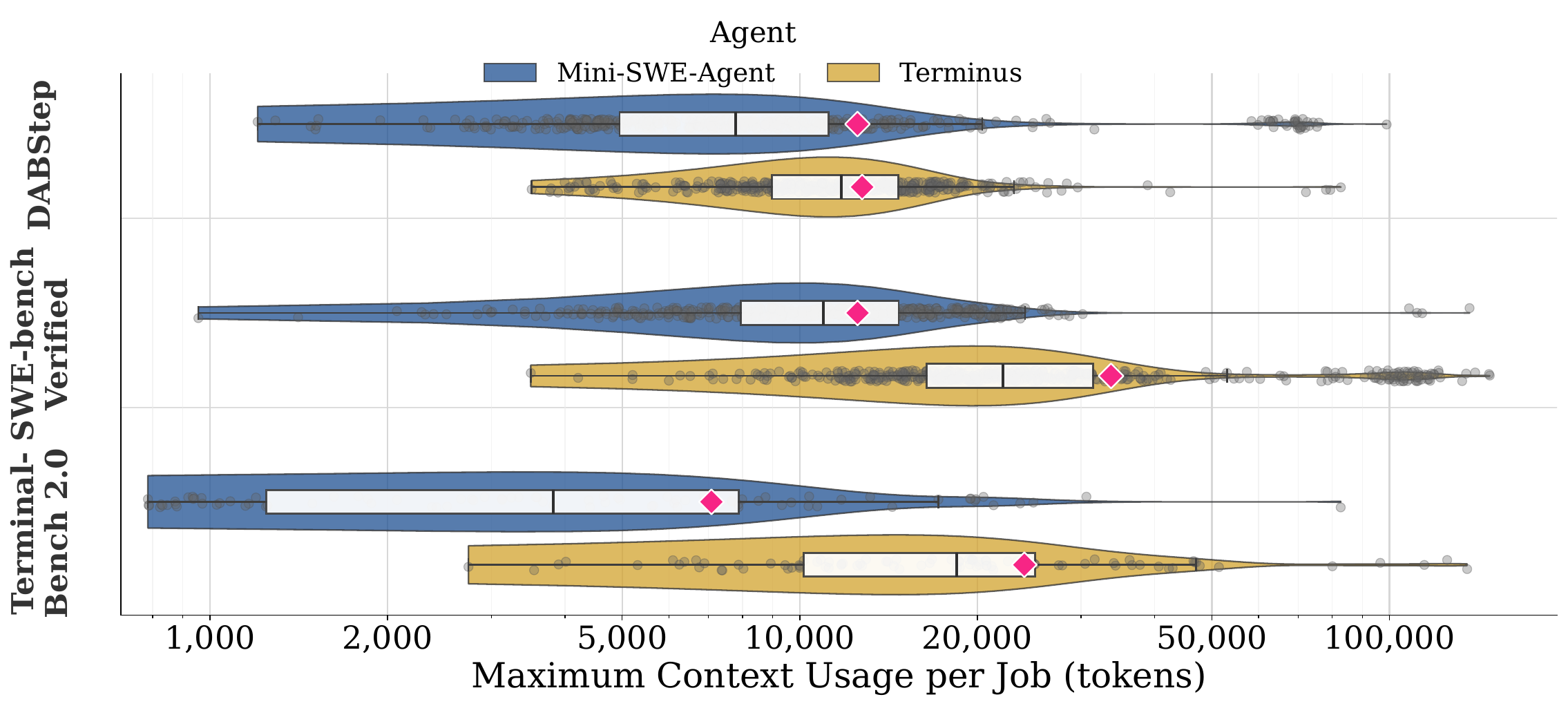}
    \caption{Distribution of maximum context length for different agents.}
    \label{fig:characterization:context_distribution}
\end{figure}
Figure~\ref{fig:characterization:context_distribution} uses the same workload setup as \S\ref{subsection:characterization:setup} and complements Figure~\ref{fig:turn_counts} by showing the distribution of the \textit{maximum context size} reached by each agent during execution.
While Figure~\ref{fig:turn_counts} characterizes variability in the number of conversation turns and total time spent in LLM execution, this figure captures the resulting peak memory footprint of each agent’s context.

The key trend is consistent across all the benchmarks.
Maximum context usage varies widely both within and across agent--dataset pairs, with a pronounced long tail of agents whose contexts grow to very large sizes.
While many agents reach peak contexts on the order of only tens of thousands of tokens, a non-trivial subset grows to hundreds of thousands of tokens.
This result reinforces that agentic serving creates highly heterogeneous and difficult-to-predict memory demand, where a small number of long-context agents can disproportionately stress the serving system.

\subsection{System Performance in Thrashing Regime} \label{appendix:threashing_performance}
\begin{figure}[t]
    \centering
    \includegraphics[width=\linewidth]{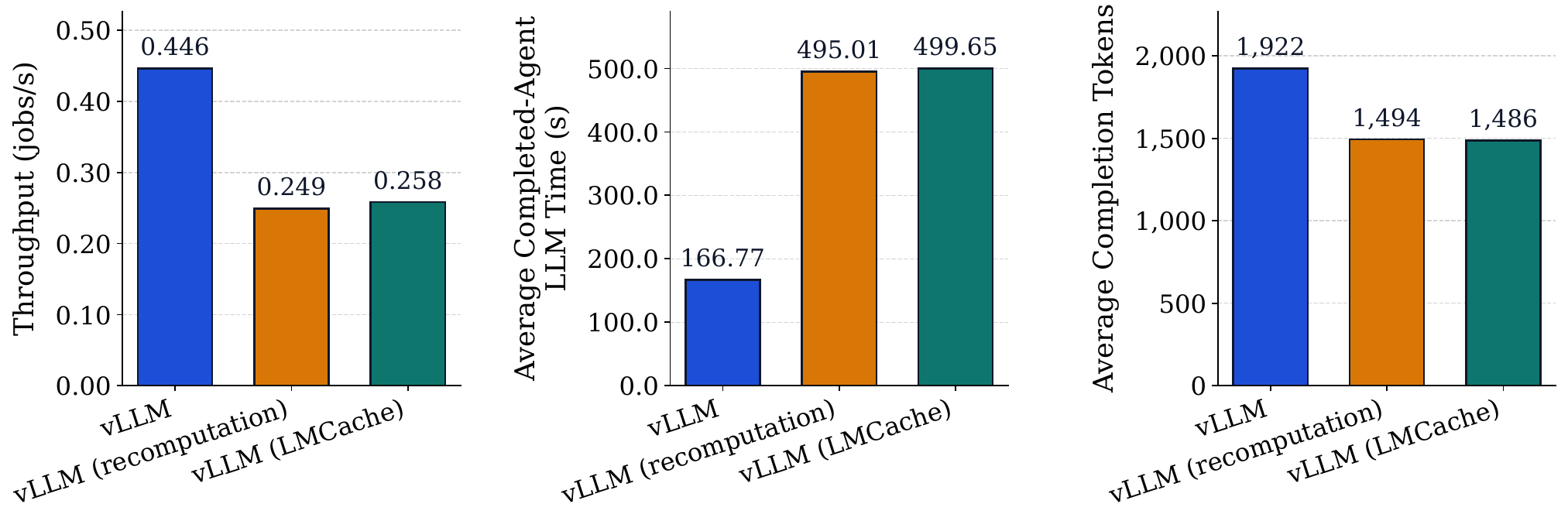}
    \caption{Average throughput (left), agent LLM time (middle), and average completion token thorughput (right) comparison of a non-thrashing vLLM baseline (0.5 jobs/s) with two thrashing baselines with recomputation and LMCache-based offloading (both 0.6 jobs/s). Across three latency and throughput metrics, the thrashing regime leads to severe performance degradation.}
\label{fig:performance_effect_of_thrashing}
\end{figure}
While operating in the thrashing regime, systems typically employ either (1) recomputation of evicted tokens~\cite{kang2026thunderagent, li2025continuum} or (2) offloading context to CPU memory or lower-tier storage using frameworks like LMCache~\cite{cheng2025lmcache}.
Figure~\ref{fig:performance_effect_of_thrashing} evaluates these two strategies (i.e., recomputation and offloading) against a baseline operating in the non-thrashing region.
To understand the effect of thrashing, we construct a controlled experiment using a short-running mini-swe-agent request from SWE-bench Verified whose conversation length is at the 25th percentile (i.e., 18 turns), and replicate this request in the input stream.
This setup simplifies regime control, while the resulting insight on the cost of thrashing is independent of the specific agent or request distribution.
We replay the same request at two arrival rates: 0.5 and 0.6 requests/second, which place the system in the non-thrashing and thrashing regimes, respectively. 
We then compare performance across the two regimes.
Our evaluation across three metrics reveals significant performance degradation for thrashing.
\begin{itemize}[leftmargin=*]
    \item \textbf{System Throughput:} Recomputation and offloading reduce average agent completion throughput by 44.2\% and 42.1\%, respectively.
    \item \textbf{LLM Latency:} The average time spent in LLM calls increases by 3$\times$ for both recomputation and offloading.
    \item \textbf{Effective Decode Throughput:} Excluding recomputation cycles, useful decode throughput drops by 22.3\%.
\end{itemize}
These results quantify the sub-optimal nature of thrashing-heavy serving.
Performance is limited by either \textit{redundant compute cycles} in the recomputation baseline or \textit{starvation of GPU compute resources} due to CPU-GPU data transfers.

\subsection{Per Agent Throughput Distribution} \label{appendix:agent_distribution}
\begin{figure}[t]
    \centering
    \includegraphics[width=\linewidth]{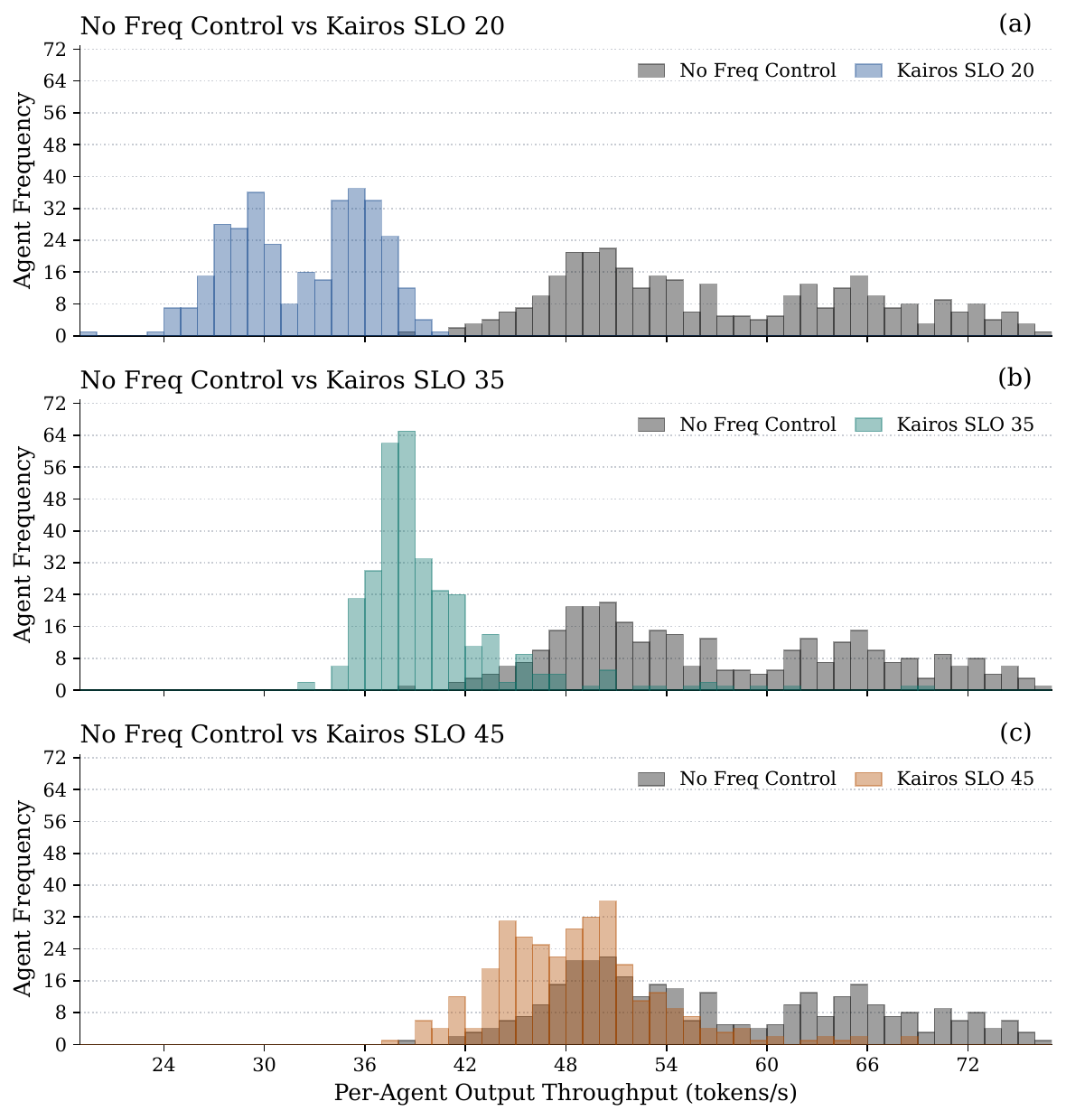}
    \caption{Per-agent throughput distribution comparing a no frequency control baseline and \THISWORK\ with different SLO targets of 20, 35, and 45 tokens/s.}
    \label{fig:agent_throughput_distribution}
\end{figure}
To explain Figure~\ref{fig:different_slo} further, Figure~\ref{fig:agent_throughput_distribution} shows the distribution of per-agent throughput for the no frequency control baseline and \THISWORK\ under different SLO targets of 20, 35, and 45 tokens/s.
Across all settings, \THISWORK\ shifts the throughput distribution left compared to the baseline, indicating that it deliberately slows down execution to reduce power. 
However, this shift is controlled: the bulk of the distribution is aligned with the respective SLO targets, showing that \THISWORK\ adapts effectively to different performance requirements rather than applying a one-size-fits-all policy.

Importantly, the figure demonstrates that \THISWORK\ slows agents just enough to meet the SLO without excessive over-provisioning.
Under loose SLOs (e.g., 20 tokens/s), throughput is tightly clustered near the target, maximizing power savings by avoiding unnecessary high-frequency execution. 
As the SLO tightens (35 and 45 tokens/s), the distribution shifts right accordingly, reflecting higher frequencies to meet stricter performance demands.
This behavior highlights that \THISWORK\ achieves its core objective: it dynamically balances power and performance by reducing excess throughput while still maintaining SLO attainment across varying targets.

\subsection{\THISWORK\ ID Tracker Interface Example} \label{app:id-tracker}

We illustrate how \THISWORK\ integrates with existing agent frameworks through a minimal, drop-in launch interface in Figure \ref{lst:id-tracker-launch}.

\begin{figure}[t]
\caption{Launching \texttt{id-tracker} with Harbor and a context-aware router}
\label{lst:id-tracker-launch}
\begin{lstlisting}[language=bash]
# `id-tracker` wraps an ordinary Harbor launch.
# It inherits the parent environment, 
# reads the per-agent API token from env,
# generates an agent name internally, 
# and forwards requests through the
# ctx-aware-router.

CTX_AWARE_ROUTER_URL="http://127.0.0.1:24157" \
OPENAI_BASE_URL="http://127.0.0.1:24157/v1" \
OPENAI_API_KEY=YOUR_KEY \
python -m id_tracker \
  --router-url "${CTX_AWARE_ROUTER_URL}" \
  -- \
  harbor run \
    --agent mini-swe-agent \
    --model \
        hosted_vllm/Qwen3-Coder-30B-A3B-Instruct \
    --dataset terminal-bench@2.0
\end{lstlisting}
\end{figure}



\section{Discussion} \label{section:discussion}
This section discusses two design and evaluation choices in detail.

\subsection{PD Disaggregation} \label{appendix:pd_disaggregation}
In this work, we characterize the complex interplay between frequency/power control, serving system performance, and context usage for serving agentic LLM workloads.
This is a common new problem faced by different serving implementation like the PD aggregated/disaggregated instances, both need to carefully deal with the persistent agent context across different requests.
In these settings, context becomes a first-class resource, and its dynamic usage directly shapes both system throughput and power-performance tradeoffs.
While our design is instantiated on a PD-aggregated architecture, the underlying challenge is fundamentally driven by context dynamics rather than a specific execution model.
We leave a full exploration of these techniques in PD-disaggregated settings to future work, where additional coordination overheads may arise.

\subsection{Choice of Request Arrival Rate} \label{appendix:arrival_rate_choice}
In real-world deployments, a given hardware configuration can sustain only a limited input request arrival rate before performance degrades.
As demand increases beyond this capacity, data center operators must scale up or scale out by using either more powerful hardware or higher number of hardware instances to maintain throughput and SLO.
Therefore, in this work, we select input arrival rates that are commensurate with the capabilities of our experimental hardware, ensuring that the system operates in a realistic and meaningful regime.
While higher arrival rates can be evaluated, doing so would require proportionally more GPU resources.
Our goal is to study power--performance trade-offs under representative operating conditions for a fixed hardware setup.

\clearpage
\section{Artifact Appendix}

\subsection{Abstract}

This artifact accompanies our paper ``KAIROS: Stateful, Context-Aware
Power-Efficient Agentic Inference Serving.''  It contains the KAIROS implementation, post-processing tools, and plotting scripts.  
The supported reviewer workflow replays recorded Terminal-Bench~2.0 and DABStep agent conversations on Qwen3-Coder-30B FP8; 
Six experiments reproduce the Figure~7 comparison among no frequency control, fixed 810-MHz frequency, and KAIROS, and the Figures~10--11 comparison showing that context-aware admission prevents thrashing.

\subsection{Artifact check-list (meta-information)}

{\small
\begin{itemize}
  \item \textbf{Algorithm:} Instance-level GPU DVFS; context-aware admission
        control; deterministic agent-conversation replay.
  \item \textbf{Program:} Python~3.12, Bash, Docker Compose, vLLM, KAIROS
        gateways and controllers, Zeus, and post-processing/plotting tools.
  \item \textbf{Compilation:} Not required; Python dependencies are locked by
        \texttt{uv.lock}.
  \item \textbf{Models:} Qwen/Qwen3-Coder-30B-A3B-Instruct-FP8 for the
        reviewer flow.
  \item \textbf{Data set:} Four recorded replay plans derived from
        SWE-Bench Verified, DABStep, and Terminal-Bench~2.0 agent runs.
  \item \textbf{Run-time environment:} Linux, Docker with Compose, NVIDIA
        Container Toolkit, an NVIDIA driver, and a running \texttt{zeusd}.
  \item \textbf{Hardware:} One NVIDIA H100 with sudo access for the six reviewer
        experiments
  \item \textbf{Run-time state:} Model weights, the published container image,
        verified replay plans, an available GPU/port profile, and a writable
        \texttt{/var/run/zeusd.sock}.
  \item \textbf{Execution:} Generate fixed experiment bundles, replay the
        conversations, post-process logs, report metrics, and render figures.
  \item \textbf{Metrics:} P5 per-agent output throughput, job throughput, GPU
        power, energy per completed agent, context occupancy, and latency.
  \item \textbf{Output:} Machine-readable summaries, tabular reports, and
        PDFs corresponding to the evaluated paper figures.
  \item \textbf{Experiments:} Six reviewer experiments for Figure~7 and
        Figures~10--11;
  \item \textbf{Disk space required:} 100GB
  \item \textbf{Time to prepare workflow:} Typically tens of minutes,
        depending on network and storage speed, excluding model and container
        downloads.
  \item \textbf{Time to complete experiments:} Approximately 18 H100 GPU hours of
        measured replay time for the six reviewer runs, plus setup
        and post-processing
  \item \textbf{Publicly available:} Yes; code is on GitHub and replay data is
        on Figshare.
  \item \textbf{Workflow automation:} Bash and Python scripts, \texttt{uv},
        and Docker Compose.
          \item \textbf{Archived:} Code DOI:
  \url{https://doi.org/10.6084/m9.figshare.34184400} ; data DOI:
        \url{https://doi.org/10.6084/m9.figshare.33236611.v2}.
\end{itemize}
}

\subsection{Description}

\subsubsection{How to access}


The source code is publicly available at
\url{https://doi.org/10.6084/m9.figshare.34184400}, and the replay plans
are available at
\url{https://doi.org/10.6084/m9.figshare.33236611.v2}. 
The entry points are \texttt{README.data.md} for required data preparation,
\texttt{README.ae.md} for the supported reviewer workflow.

\subsubsection{Hardware dependencies}

The reviewer workflow requires one NVIDIA H100 NVL with sudo access.  
GPU frequency control requires access to a \texttt{zeusd} daemon through \texttt{/var/run/zeusd.sock}.

\subsubsection{Software dependencies}

The host requires Linux, Python~3.12, \texttt{uv}, Docker with Compose, NVIDIA
Container Toolkit, \texttt{curl}, \texttt{tar}, \texttt{md5sum}, and
\texttt{sha256sum}.  Python packages are pinned in \texttt{uv.lock}.  Serving
uses the published image
\texttt{yichaoyuan/vllm-vllm-openai:v0.14.0-otel-lp-lmcache}, which includes
vLLM~0.14.0 and the artifact's serving extensions.  Hugging Face access is
needed if the model is not already cached.

\subsubsection{Data sets}

The version-2 Figshare archive contains exactly four replay-plan files with a
total logical size of approximately 29.8~GB.  \texttt{README.data.md} records
the immutable download, byte count, archive MD5, extraction paths, and
per-file SHA-256 verification.  The reviewer flow uses the DABStep and
Terminal-Bench plans.

\subsubsection{Models}

All reviewer experiments use Qwen/Qwen3-Coder-30B-A3B-Instruct-FP8. 

\subsection{Installation}

Clone the code repository and enter its root.  First follow
\texttt{README.data.md} to download, verify, and extract the replay plans.
Then run \texttt{uv sync {-}{-}frozen} and
\texttt{uv run python ae/check\_env.py {-}{-}gpu 0 {-}{-}port-profile 0}.
The optional \texttt{{-}{-}verify-checksums} flag performs the complete
SHA-256 validation through the environment checker.  If the administrator
does not provide \texttt{zeusd}, start it as described in
\texttt{README.ae.md}; set the Hugging Face cache and token variables only
when model download is required.

\subsection{Experiment workflow}

\begin{enumerate}
  \item Complete the download, extraction, and four SHA-256 checks in
        \texttt{README.data.md}.
  \item Generate the six exact bundles with
        \texttt{uv run python ae/generate\_experiments.py {-}{-}gpu 0
        {-}{-}port-profile 0}.
  \item Run the six keys listed in \texttt{README.ae.md} sequentially with
        \texttt{uv run python ae/run\_experiment.py KEY}.  Each command starts
        the matching vLLM/gateway stack, performs one three-hour replay, and
        tears the stack down.
  \item Run \texttt{uv run python ae/postprocess.py} followed by
        \texttt{uv run python ae/report.py}.  Materialize and plot
        Figures~10 and~11 using the commands in \texttt{README.ae.md}.
\end{enumerate}

\subsection{Evaluation and expected results}
For the six-experiment AE workflow, the reproduced Terminal-Bench result from Figure~7 and the DABStep results from Figures~10 and~11 should match the corresponding results reported in the paper, allowing for small differences due to host and GPU measurement noise.

\end{document}